\documentclass[12pt,prb,aps]{revtex4}
\usepackage{epsf}
\usepackage{epsfig}
\begin{document}
\title{Stable Equilibrium Based on L\'evy Statistics: A Linear Boltzmann Equation Approach}
\author{Eli Barkai}
\affiliation{Dept. of Chemistry and Biochemistry, Notre Dame University, Notre Dame, IN 46556.}
\email{jbarkai@nd.edu}
\date{\today}

\begin{abstract}

To obtain further insight on possible power law generalizations
of Boltzmann equilibrium concepts, a stochastic 
collision model is investigated.
We consider
the dynamics
of a tracer particle of mass $M$,  undergoing elastic collisions 
with ideal gas particles of mass $m$, in the Rayleigh limit $m<<M$. 
The probability density function (PDF) of the gas particle velocity
is $f\left( \tilde{v}_m \right)$.
Assuming a uniform collision rate and molecular chaos,
we obtain the equilibrium distribution
for the velocity of the tracer particle $W_{eq}(V_M)$.
Depending on
asymptotic properties of $f\left( \tilde{v}_m \right)$ we find that
$W_{eq}(V_M)$ is either
the Maxwell velocity distribution or a L\'evy  
distribution. 
In particular our results yield 
a generalized Maxwell distribution based on L\'evy 
statistics using two  approaches.
In the first a thermodynamic argument is used, imposing on the dynamics
the condition
that equilibrium properties of the heavy tracer particle be independent
of the coupling $\epsilon=m/M$ to  the gas particles,
similar to what is found for a Brownian particle in a fluid.
This approach leads to a generalized temperature concept.
In the second approach it is assumed that bath particles velocity
PDF scales with an energy scale, i.e. the (nearly) ordinary temperature,
as found in standard statistical
mechanics. The two approaches yield different types of L\'evy
equilibrium which merge into a unique solution only for
the Maxwell--Boltzmann case. 
Thus, relation between thermodynamics and statistical mechanics
becomes non-trivial for the power law case.
Finally, the relation of the kinetic model to fractional
Fokker--Planck equations is  discussed.

\end{abstract}

\maketitle

\section{Introduction}

 Khintchine \cite{Khin} 
revealed the deep relation between
the Gaussian 
central limit theorem and classical Boltzmann--
Gibbs statistics. From a simple stochastic point of view, we may
see this relation by considering the velocity of a
Brownian particle. 
The most basic phenomenological dynamical description
of Brownian motion, is in terms of the Langevin 
equation
$\dot{V}_M= - \gamma V_M + \eta(t)$, where $\eta(t)$ is a {\em  Gaussian} white
noise term.
Usually the assumption of Gaussian noise is imposed on the Langevin
equation to obtain a Maxwellian velocity distribution
describing the equilibrium of the Brownian particle. 
We may reverse our
thinking of the problem, the Gaussian noise is naturally expected 
based on central limit theorem arguments, and the latter
leads to Maxwell's equilibrium. Similar arguments
hold for a Brownian particle in an external time
independent binding force field, the Maxwell--Boltzmann
equilibrium is obtained in the long time limit
only if the noise term is Gaussian.  
Yet another case where the relation between Gaussian
statistics and statistical mechanics is seen is
Feynman's Gaussian path integral formulation of
statistical mechanics. 

 However, Gaussian central limit theorem is non unique.
L\'evy, and Khintchine \cite{Feller} have generalized the
Gaussian central limit theorem, to the case of summation
of independent, identically distributed random variables described
by long tailed 
distributions. In this case L\'evy distributions replace 
the Gaussian in generalized limit
theorems.  Hence it is natural to ask 
\cite{Montroll,Grigolini,Chechkin,Zannette}
if a L\'evy
based statistical mechanics exist? And if so what is
its  physical domain and its relation to thermodynamics.
This type of questions are timely due to the interest
in generalizations of statistical mechanics 
\cite{Abe,Long,CohenEDG,Gallavotti}, mainly non-extensive statistical
mechanics due to Tsallis \cite{Tsallis}.
While L\'evy statistics is used in many
applications \cite{Bouch1,review,Cohen,Zaslavsky,Barkai1,Barkai2,Barkai3},
its {\em possible} relation
to generalized
equilibrium statistical mechanics is still unclear. 

 As well known, Boltzmann used a kinetic approach
for a dilute gas of particles to derive Maxwell's
velocity distribution, his starting point being 
his non-linear equation (see conditions in
\cite{Cercignani}). Thus, the kinetic approach can be used as
a tool to derive equilibrium, starting from non-equilibrium
dynamics. Ernst \cite{Ernst}, shows how the Maxwell equilibrium
is obtained from several stochastic collision models.
More recently, Chernov and Lebowitz \cite{CL} investigated numerically
a gas $+$ heavy piston system,
showing that the gas particle velocity distribution
approaches a Maxwellian 
distribution, starting from a non-Maxwellian initial condition.
Thus, the Maxwell equilibrium transcends details of individual 
kinetic models. It is important to note
that in these works, important boundary 
and initial conditions are imposed on
the dynamics \cite{Cercignani,Ernst,CL,Lar}. 
For example, the second moment of the
velocity of the particles is supposed to be finite,
Eqs. 1.7 and 2.10 in \cite{Ernst}.
A possible domain of power law generalizations
of Maxwell's equilibrium, are the cases where
one does not impose `finite variance' initial and/or
boundary conditions.

 In this manuscript we will consider a simple kinetic approach to obtain a 
generalization of Maxwell's velocity distribution.
Briefly we consider a one dimensional tracer particle of mass $M$
randomly colliding with gas particles of mass $m<<M$. 
Two main assumptions are used: (i) molecular chaos holds,
implying lack of correlations in
the collision  process
(Stoszzahlansantz),
and (ii)  rate of collisions is independent of the
energy of the colliding particles. 
 Let the probability
density function (PDF) of
velocity of
the gas particles be 
$f\left( \tilde{v}_m \right)$.
If $f\left( \tilde{v}_m \right)$ is Maxwellian, the model
describes Brownian type of motion for the heavy particle,
the equilibrium being the Maxwell distribution \cite{Kampen,Opp,BF}. 
Since our aim is to investigate generalized equilibrium,
we do not impose the standard condition of Maxwellian
velocity distribution for the bath particles.
 Our aim is to show, under general conditions,
that the equilibrium PDF of the tracer particle
$W(V_M)$ is stable. Stable means here that large
classes of velocity distributions for bath particles $f( \tilde{v}_m)$
will yield a unique type of equilibrium for the tracer particle
$W_{eq}(V_M)$. Namely the equilibrium velocity distribution
transcends the details of precise shape of $f(\tilde{v}_m)$.
One of those stable velocity distributions
will turn out to be 
Maxwell's velocity distribution.

 To obtain an equilibrium we use two approaches.
The first we call the thermodynamic approach, since
it is based on the following thermodynamic argument.
We know that two extensive thermal systems, A and B,
left at thermal contact will obtain a thermal
equilibrium with respect to each other. While the interaction
between the two systems is responsible for the
equilibration, the precise nature of this interaction
is not important for the equilibrium properties of the 
composite system. In particular statistical 
properties of system A are independent of the coupling
parameters between the two systems. The idea is that
interface interaction energy is much smaller than the total energy
of the two systems, and hence may be neglected.
For a Brownian particle (system A) in a liquid (system B), this 
implies that the equilibrium velocity distribution of the Brownian particle
is independent of the physical and chemical nature of the liquid (besides
its temperature).
In our model the coupling between the tracer particle
and bath particle is the mass ratio $\epsilon = m/M$.
In the thermodynamic approach we will impose on the dynamics,
the condition that the equilibrium of the
heavy tracer particle  is independent of the coupling $\epsilon$.
This type of dynamics yields a generalized temperature concept
$T_{\alpha}$.

 In the second approach, we use a scaling argument based on
the assumption that an energy scale controls the velocity
distribution of the bath particles. This means
that we assume
\begin{equation}
 f\left( \tilde{v}_m \right) = {1 \over \sqrt{T/m}} q(\tilde{v}_m/\sqrt{T/m}),
\label{eq00}
\end{equation}
and $q(x)$ is a non-negative normalized function.
Here $T$  has similar meaning as the  
usual temperature.
From this general starting point stable equilibrium is 
derived for the tracer particle.

 It is shown that the two approaches yield  stable
L\'evy equilibrium. However, besides the Maxwellian case
the two approaches yield different types of equilibrium.  Thus, according
to our model generalized equilibrium based on
L\'evy statistics naturally emerges, however this
type of equilibrium is different from the standard
equilibrium. One cannot generally treat the tracer particle
equilibrium properties as separable from
coupling to the bath, and at the same time 
use standard temperature concept $T$.

 This manuscript is organizes as follows. In Sec. \ref{SecModel}
I present the model, and the linear Boltzmann equation
under investigation.  The time dependent solution of the
model is found in Fourier space. In Sec. 
\ref{SecEquil} the equilibrium solution of the Boltzmann
equation is obtained, this solution is valid for any
mass ratio $\epsilon$. We then consider the limit $\epsilon <<1$,
using both the thermodynamic approach 
Sec. \ref{SecTher} and scaling approach Sec. 
\ref{SecSca}. Throughout this work numerically 
exact solutions of the model are compared with asymptotic 
solution obtained in the limit of weak collisions.
We then turn back to dynamics, in Sec. 
\ref{SecBack}
we discuss
the relation of the model with 
fractional Fokker--Planck
equations \cite{review,Zaslavsky,SokKB}.
We end with a brief summary.

\section{Model and Time Dependent Solution}
\label{SecModel}

 We consider a one dimensional tracer particle with the mass $M$
coupled with bath particles of mass $m$. The tracer particle velocity
is $V_M$. At random times the tracer particle collides
with bath particles whose velocity is denoted with $\tilde{v}_m$.
Collisions are elastic hence from conservation of momentum and energy
\begin{equation}
V_M ^{+} = \xi_1 V_M ^{-} + \xi_2 \tilde{v}_m,
\label{eq01}
\end{equation}
where 
\begin{equation}
\xi_1 = { 1 - \epsilon \over 1 + \epsilon} \ \ \ \ \xi_2 = { 2 \epsilon \over 1 + \epsilon}
\label{eq02}
\end{equation}
and $\epsilon \equiv  m/M$ is the mass ratio. In Eq. (\ref{eq01}) $V_M ^{+}$
$(V_M ^{-})$
is the velocity of the tracer particle after (before) a collision event. 
The duration of the collision events is much shorter than
any other time scale in the problem.
The collisions occur at a uniform rate $R$ independent of the velocities
of colliding particles. The probability density function (PDF)
 of the bath particle
velocity is $f(\tilde{v}_m)$. This PDF does not change during the collision
process, indicating that re-collisions of the bath 
particles and the tracer particle are neglected.
Note that if velocities of bath and tracer particle
are identical before collision event $\tilde{v}_m = V_M ^-$,
 we have $V_M ^{+} = V_M ^-$,
a fact worth mentioning since two particle with identical velocities
never collide in one dimension. 

 We now consider the equation of motion for the tracer particle
velocity PDF $W(V_M , t)$ with initial conditions concentrated on
$V_M(0)$. Standard kinetic considerations yield
\begin{widetext}
\begin{equation}
 {\partial W\left(V_M ,t \right) \over \partial t} = 
 - R W\left( V_M, T \right)
 + R \int_{-\infty}^{\infty} {\rm d} V_M ^{-} \int_{-\infty}^{\infty} {\rm d} \tilde{v}_m  
W \left(V_M ^{-}, t\right)f\left( \tilde{v}_m \right) \times 
\delta\left( V_M - \xi_1 V_M ^{-} - \xi_2 \tilde{v}_m \right),
\label{eq03}
\end{equation}
\end{widetext}
where the delta function gives the constrain on energy and
momentum conservation in collision events. Us-usual the first (second) term in
Eq. (\ref{eq03}) describes a  tracer particle leaving (entering)
the velocity point $V_M$ at time $t$.
Eq. (\ref{eq03}) yields the forward master equation,
also called the linear Boltzmann equation 
%
\begin{equation}
{\partial W\left(V_M ,t \right) \over \partial t} =   
- R W\left( V, T \right) +{ R \over \xi_1} \int_{-\infty}^{\infty} {\rm d} \tilde{v}_m 
W \left( { V_M - \xi_2 \tilde{v}_m \over \xi_1} \right) f \left( \tilde{v}_m \right).
\label{eq03a}
\end{equation}
%
This equation is valid for $\xi_1 \ne 0$ namely $\epsilon \ne 1$.
In Eq. (\ref{eq03a}) the second term on the right hand side is
a convolution in the velocity variables, hence
we will consider the problem in Fourier space.
Let
$\bar{W}(k,t)$
 be the Fourier transform of the velocity PDF 
\begin{equation}
\bar{W} \left( k , t \right) = \int_{-\infty}^{\infty} W\left( V_M , t \right) \exp\left( i k V_M \right) {\rm d} V_M,
\label{eq04}
\end{equation}
we call $\bar{W}(k,t)$ the tracer particle characteristic function.
Using Eq. (\ref{eq03a}), the equation of motion for 
$\bar{W} \left( k , t \right)$  is a finite difference
equation
\begin{equation}
{ \partial \bar{W}(k,t) \over \partial  t} = - R \bar{W}(k,t) +
R \bar{W} \left( k \xi_1 , t \right) \bar{f} \left(k \xi_2  \right),
\label{eq05}
\end{equation}
where $\bar{f} \left( k \right) $ is the Fourier transform of
$f(\tilde{v}_m)$. 
 In Appendix A 
the solution 
of the equation of motion Eq. (\ref{eq05}) is obtained by iterations
\begin{equation}
\bar{W}\left( k, t \right) = 
\sum_{n=0}^{\infty} { \left( R t \right)^n \exp\left( - R t \right)  \over n!} e^{ i k V_M(0) \xi_1^n } \Pi_{i=1}^n \bar{f}\left( k \xi_1^{n-i} \xi_2 \right),
\label{eq06}
\end{equation}
with the initial condition  $\bar{W}( k , 0 ) = \exp[ i k V_M (0)]$.
Similar analysis for the case $\xi_1=0$ shows that Eq. (\ref{eq06})
is still valid with 
$\bar{f}\left(k \xi_1^{n-i} \xi_2 \right)=\bar{f}( k ) \delta_{ni}$ and
$\xi_1^n=\delta_{n0}$ where $\delta_{ni}$ is the Kronecker symbol.

The solution Eq. (\ref{eq06}) has a simple interpretation.
The probability that the tracer particle has collided $n$ times with
the bath particles is given according to the Poisson law
\begin{equation}
 P_n(t)= { \left( R t  \right)^n \over n!} \exp\left( - R t \right),
\label{eq07}
\end{equation}
reflecting the assumption of uniform collision rate.
Let $W_n(V_M)$ be the PDF of the tracer particle conditioned that
the particle experiences $n$ collision events. It can be shown
that the Fourier transform of
$W_n(V_M)$ is
\begin{equation}
\bar{W}_n(k) = 
 e^{ i k V_M(0) \xi_1^n } \Pi_{i=1}^n \bar{f}\left( k \xi_1^{n-i} \xi_2 \right).
\label{eq07aa}
\end{equation}
Thus Eq. (\ref{eq06}) is a sum over the probability of having $n$
collision events in time interval $(0,t)$ times the Fourier transform
of the velocity PDF after exactly $n$ collision event
\begin{equation}
\bar{W}(k,t) =  \sum_{n=0}^{\infty} P_n(t) \bar{W}_n(k).
\label{eq08}
\end{equation}
It follows immediately that the solution of the problem is
\begin{equation}
W(V_M ,t) =  \sum_{n=0}^{\infty} P_n(t) W_n(V_M),
\label{eq09}
\end{equation}
where $W_n(V_M)$ is the inverse Fourier transform of $\bar{W}_n(k)$
Eq. (\ref{eq07aa}).


{\bf Remark 1} The history of the model and its relatives
for the case when
$f(\tilde{v}_m)$ is Maxwellian is long. Rayleigh, who wanted
to obtain insight into the Boltzmann equation, investigated the
limit $\epsilon \ll 1$. This important limit describes dynamics of a heavy
Brownian particle in a bath of light gas particles,
according to the Rayleigh equation \cite{Kampen}. 
The limit $\epsilon = 1$, when collisions are impulsive, 
is called the 
Bhantnagar--Gross--Krook limit.
 More recent work considers
this model for the case where an external field is acting on the
tracer particle, for example in the context of calculation
of activation rates over a potential barrier \cite{BF,Berez}. 
In the Rayleigh limit
of $\epsilon \to 0$ one obtains the dynamics of the Kramers
equation, describing Brownian motion in external force field. 
Investigation of the model for the case where
collisions follow a general renewal process (i.e.,
non Poissonian) was considered in \cite{BS}.
Chernov, Lebowitz and Sinai \cite{Sinai} 
used
a rigorous approach for a related model, 
one of the aims being  
the investigation of the validity of hydrodynamic equations
of motion for scaled coordinates of the heavy tracer particle.

\subsection{ An Example:- L\'evy Stable Bath Particle Velocities}

 In the classical works on Brownian motion the
condition that bath particle velocity distribution is
Maxwellian is imposed. As a result one obtains an equilibrium
 Maxwellian  
distribution for the tracer particle (i.e., detailed balance is imposed
on the dynamics).
This behavior is not unique to Gaussian process, in the sense 
that if we choose a L\'evy stable law to describe the
bath particle velocity PDF, the tracer particle 
will obtain an equilibrium which is also a L\'evy distribution.
This property does not generally hold
for other choices of bath particle velocity PDFs.

  To see this let the PDF of bath particle velocities
be a symmetric L\'evy density, in Fourier space
\begin{equation}
\bar{f} \left( k \right) = \exp\left[ - {A_{\alpha} | k|^{\alpha}\over \Gamma(1 + \alpha) } \right],
\label{eq10}
\end{equation}
and $0<\alpha \le 2$. The special case $\alpha=2$ corresponding to the
Gaussian PDF. We will 
discuss later the dependence of the parameter
$A_{\alpha}$ in Eq. (\ref{eq10}) on mass of bath particles $m$
and on a generalized temperature concept. Using Eqs. 
(\ref{eq06},\ref{eq10})
we obtain   
\begin{equation}
  W\left( V_M , t \right) = 
\sum_{n=0}^{\infty} { e^{- R t} \left( R t \right)^n \over n!} { 1  \over \left[ A g_{\alpha} ^n ( \epsilon) \right]^{1/\alpha} }
l_{\alpha} \left\{ { \left[ V_M - V_M\left(0\right) \xi_1^n \right] \over \left[ A g_{\alpha} ^n \left( \epsilon \right) \right]^{1/\alpha} } \right\},
\label{eq11}
\end{equation}
where $l_{\alpha}(x)$ is the symmetric L\'evy density
whose Fourier pair is 
\begin{equation}
\bar{l}_{\alpha}(k) = \exp( - |k|^{\alpha}),
\label{eq11a}
\end{equation}
$A=A_{\alpha}/\Gamma(1 + \alpha)$,
and
\begin{equation}
g_{\alpha} ^n \left( \epsilon \right) \equiv \xi_2^{\alpha} { 1 - \xi_1 ^{ \alpha n } \over 1 - \xi_1 ^{\alpha} }.
\label{eq12aa}
\end{equation}
Later we will use the $n \to \infty$ limit of Eq. (\ref{eq12aa})
\begin{equation}
g_{\alpha} ^\infty \left( \epsilon \right) = {\left( 2 \epsilon\right)^{\alpha} \over \left( 1 + \epsilon\right)^\alpha - \left( 1 - \epsilon\right)^\alpha },
\label{eq12a}
\end{equation}
and the small $\epsilon$ behavior 
\begin{equation}
g_{\alpha} ^\infty (\epsilon ) \sim {2^{\alpha -1} \over \alpha} \epsilon^{\alpha -1}.
\label{eq12b}
\end{equation}
From Eq. (\ref{eq11})
we see that for all times $t$ and for any mass ratio $\epsilon$, the
tracer particle velocity PDF is a sum of rescaled bath particle
velocities PDFs. 
In the limit $t \to \infty$ a stationary state is
reached
\begin{equation}
 W_{eq}\left(V_M \right) = 
{ 1 \over \left[ A g_{\alpha} ^\infty ( \epsilon) \right]^{1/\alpha} }
l_{\alpha} \left\{ { V_M  \over \left[ A g_{\alpha} ^\infty \left( \epsilon \right) \right]^{1/\alpha} } \right\},
\label{eq11bb}
\end{equation}
or in Fourier space 
\begin{equation}
\bar{W}_{eq}\left(k \right) = 
\exp\left[ - A g_{\alpha} ^\infty \left( \epsilon \right) |k|^{\alpha} \right].
\label{eq12}
\end{equation}
Thus  the distribution of $\tilde{v}_m$ and $V_M$ differ only
by a scale parameter. For non--L\'evy PDFs of bath particles velocities
this is not the general case, the distribution of $V_M$ 
differs from that of $\tilde{v}_m$.

\section{Equilibrium}
\label{SecEquil}

 In the long time limit, $t \to \infty$ the tracer particle characteristic
function 
reaches an equilibrium
\begin{equation}
\bar{W}_{eq}(k) \equiv \lim_{t \to \infty} \bar{W}(k, t ).
\label{eq13zzz}
\end{equation}
 This equilibrium is  obtained from Eq. (\ref{eq06}). We notice that when
$R t \to \infty$,  $P_n(t) = (R t)^n \exp(- R t)/n!$ is peaked
in the vicinity of $\langle n \rangle = R t$ hence it 
is easy to see that
\begin{equation}
\bar{W}_{eq}\left( k\right) = \lim_{n\to \infty}\Pi_{i=1}^n \bar{f}\left( k \xi_1^{n-i} \xi_2 \right).
\label{eq13}
\end{equation}
In what follows  we  investigate properties of
this equilibrium.
 
 We will consider the weak collision limit $\epsilon \to 0$. 
This limit is important since number of collisions needed for
the tracer particle to reach an equilibrium is very 
large. Hence in this case we may expect the emergence of
a general equilibrium concept which is not sensitive to the
precise details of the velocity  PDF $f(\tilde{v}_m )$ 
of the bath particles. In this limit we may also
expect that in a statistical sense $\tilde{v}_m \ll V_M$, hence
the assumption of a uniform collision rate is reasonable in this
limit.  


 As mentioned in the introduction we will use two approaches to
determine the equilibrium distribution. The first is based
on a thermodynamic argument. 
We will demand that
equilibrium properties of the heavy tracer particle are
independent of the mass ratio $\epsilon$, which is the coupling
between the bath and the heavy particle in our model.

 The second approach is based
on a scaling argument. 
We assume that statistical properties of bath particles
velocities can be characterized with an energy scale $T$,
which will turn out to be the usual temperature. 
This approach is based on the assumption that an energy scale
determines the equilibrium properties of the system, similar to
what is found in ordinary statistical mechanics. In contrast for the
thermodynamic approach, the statistical measure is a generalized
temperature $T_{\alpha}$ whose units are
$\mbox{Kg}^{\alpha -1} \mbox{Mt}^{\alpha} / \mbox{Sec}^{\alpha} $.

\begin{figure}[htb]
\epsfxsize=20\baselineskip
\centerline{\vbox{
        \epsfig{file=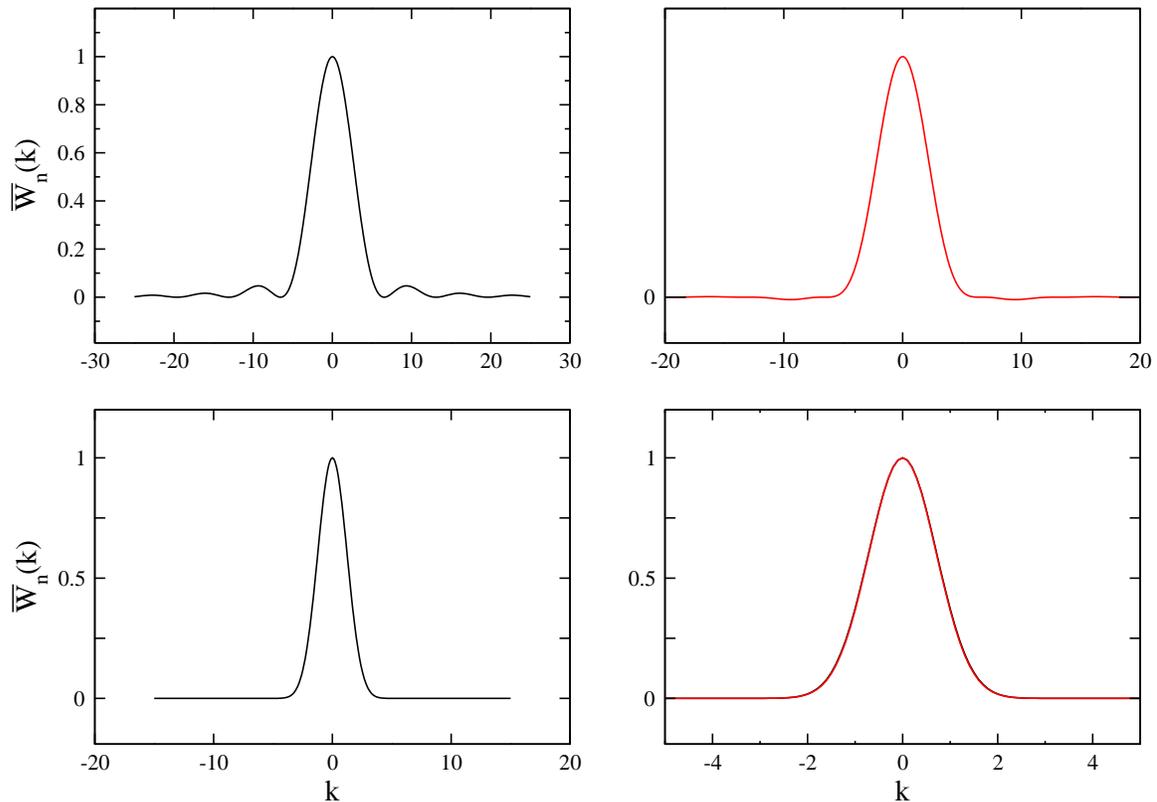, width=0.85\linewidth, angle=-90}  }}
\caption {
We show the dynamics of the collision process: 
the tracer particle characteristic function, conditioned that
exactly $n$ collision events have occurred,  $\bar{W}_n(k)$ versus $k$.
The velocity PDF of the bath particle is uniform
and $\epsilon=0.01$. We show $n=1$  (top left), $n=3$
(top right), $n=10$ (bottom left) and $n=100$ $n=1000$ (bottom right).
For the latter case we have $\bar{W}_{100}(k) \simeq \bar{W}_{1000}(k)$,
hence the process has roughly converged after $100$
collision events. The equilibrium is well approximated with a
Gaussian characteristic function
 indicating that a Maxwell--Boltzmann equilibrium is
obtained. 
}
\label{fig0}
\end{figure}

\newpage

{\bf Remark 1}  
According to Eq. (\ref{eq07aa}),
after a single collision
event the PDF of the tracer particle in Fourier space
is 
$\bar{W}_{1}(k) = 
\bar{f}\left( k \xi_2 \right)$ provided that
$V_M(0)=0$. After the second collision
event 
$\bar{W}_{2}(k) = 
\bar{f}\left( k \xi_1 \xi_2 \right)
\bar{f}\left( k \xi_2 \right)$
and after $n$ collision events
\begin{equation}
\bar{W}_{n}(k) = 
\Pi_{i=1}^n \bar{f}\left( k \xi_1^{n-i} \xi_2 \right).
\label{eqPro}
\end{equation}
This process is described in Fig. 
\ref{fig0}, where we show $\bar{W}_n(k)$ for $n=1,3,10,100,1000$.
 In this example we use 
a uniform distribution of the bath particles PDF  
Eq. (\ref{EqNE03}),
with $\epsilon =  0.01$, and $T=1$.
After roughly $100$ collision events 
the characteristic function $\bar{W}_n(k)$ reaches a
stationary state, which as we will show is well approximated by
a Gaussian (i.e., Maxwell--Boltzmann velocity PDF is obtained).
Such an equilibrium is not specific to the initial choice
of $\bar{f}(k)$, i.e., the assumption of uniform
distribution of bath particle velocities. 
Rather one of my aims is to show that
many choices of $\bar{f}(k)$ will flow towards the
Gaussian attractor (this resembles flows towards fixed points
however now we are working in function space). As we will show these flows do
not always belong to the domain of attraction of the
Gauss--Maxwell--Boltzmann behavior.

{\bf Remark 2} The equilibrium described by Eq.
(\ref{eq13}) is valid for 
a larger class  of collision models provided that two requirements
are satisfied. To see this consider Eq.
(\ref{eq09}), this equation is clearly not limited to the model
under investigation. For example if number of collisions is described
by a renewal process, equation (\ref{eq09}) is still  valid
(however generally $P_n(t)$ is not described by the Poisson law). 
The important first requirement is that in the limit $t\to \infty$
$P_n(t)$ is peaked around $n \to \infty$, and that $P_n(t)$ is not too wide
(any renewal process with finite mean satisfies this condition).
We may expect that this is the case from standard central limit theorem,
and law of large numbers. The second requirement is that recolision of 
bath particles and tracer particle are not important. This assumption
is important since without it the simple form of 
$\bar{W}_n(k)$ is not valid, and hence also Eq. (\ref{eq13}).
Physically, this means that the bath particles  maintain their own equilibrium
throughout the collision process, namely fast relaxation to 
equilibrium of the bath. 

{\bf Remark 3} The problem of analysis of the equilibrium 
Eq. (\ref{eq13}) is different from the classical 
mathematical problem of summation of independent identically 
distributed random variables \cite{Feller}. The rescaling of $k$
seen in
Eq. (\ref{eq13}) $k \to k \xi_1^{n-i} \xi_2$,  means
that we are treating a problem of summation of independent,
though non identical random variables.  The number of these random
variables is $n$, unlike standard problems of summation
of random variables, $n$ also enters in the rescaling of $k$
as seen in   $k \xi_1^{n-i} \xi_2$.

{\bf Remark 4} If $m=M$ we find $\bar{W}_{eq}(k)=\bar{f}(k)$,
this behavior is expected since in this strong collision limit
a single collision event is needed for relaxation
of tracer particle to equilibrium. This trivial equilibrium 
is not stable in the sense that perturbing $\bar{f}(k)$ 
yields a new equilibrium for the tracer particle.

\section{Thermodynamic Approach}
\label{SecTher}

 In this section the thermodynamic approach is used, imposing
the condition 
that $\bar{W}_{eq}(k)$ is independent of the coupling constant
$\epsilon$. 

\subsection{Gauss--Maxwell--Boltzmann Equilibrium}

 We investigate the cumulant expansion of $\bar{W}_{eq}(k)$, assuming
that moments of the bath particle velocity PDF are finite.
We also assume that $f( \tilde{v}_m )$ is even
as might be expected from symmetry, and from the
requirement that the total momentum of the bath is zero.

We use the long wave length expansion
\begin{equation}
\bar{f} (k) = 1 - {k^2 \over 2} \langle \tilde{v}_m ^2 \rangle + {k^4 \over 4!} \langle \tilde{v}_m ^4 \rangle + O(k^6),
\label{eq14}
\end{equation}
where
$\langle \tilde{v}_m ^2 \rangle$
($\langle \tilde{v}_m ^4 \rangle$) are the second (forth) moment of bath
particles velocity. 
Using Eq. (\ref{eq13})
\begin{equation}
 \ln\left[\bar{W}_{eq}\left( k\right)\right]  = 
\lim_{n\to \infty} \sum_{i=1}^n 
\ln\left[  \bar{f}\left( k \xi_1^{n-i} \xi_2 \right) \right].
\label{eq15}
\end{equation}
Inserting Eq. (\ref{eq14}) in (\ref{eq15}) we obtain
\begin{equation}
  \ln\left[\bar{W}_{eq}\left( k\right)\right] =
- {\langle \tilde{v}_m ^2 \rangle \over 2} g_2 ^\infty (\epsilon) k^2  + {\langle \tilde{v}_m ^4 \rangle - 3 \langle \tilde{v}_m ^2 \rangle^2 \over 4!} g_4 ^\infty(\epsilon) k^4 + O (k^6).
\label{eq16}
\end{equation}
 Using Eq.
(\ref{eq12}) with $\alpha=2$
 we obtain in the limit $\epsilon \to 0$
\begin{equation}
  \ln\left[\bar{W}_{eq}\left( k\right)\right] =
- {\langle \tilde{v}_m ^2 \rangle \over 2} { m \over M} k^2  + {\langle \tilde{v}_m ^4 \rangle -3 \langle \tilde{v}_m ^2 \rangle^2 \over 4!}2 \left( { m \over M} \right)^3 k^4 + O\left( k^6 \right),
\label{eq17}
\end{equation}
where the dependence on mass ratio $m/M$ was explicitly included.
To obtain equilibrium using the thermodynamic approach, we
require that $\bar{W}_{eq}(k)$ be independent of the mass ratio $\epsilon$.
From this physical requirement and Eq. (\ref{eq17})
we obtain a standard relationship
\begin{equation}
\langle \tilde{v}_m ^2 \rangle = {T_2 \over m}.
\label{eq18}
\end{equation}
This requirement and Eq. (\ref{eq17})
 means that dependence of mean square velocity
of tracer particle and bath particle scale with their masses
in a similar way, i.e., like $1/m$ and $1/M$ respectively.
In Eq. (\ref{eq18}) $T_2$ is the temperature,
below we show that using similar arguments for 
power law case leads to  a generalized temperature
$T_{\alpha}$.
We further demand that the forth moment of $f(\tilde{v}_m)$
will scale with the mass of the bath particle like
\begin{equation}
\langle \tilde{v}_m ^4 \rangle= q_4  \left({  T_2 \over m} \right)^2,
\label{eq19}
\end{equation}
where $q_4$ is a dimensionless constant.
This imposed condition
results in a velocity PDF
for the tracer particle, which is independent of
$q_4$ and hence of details of bath particle
velocity PDF.
In this case the equilibrium properties of the tracer 
particle become independent of 
the details of $f\left( \tilde{v}_m \right)$ (e.g., independent
of $q_4$), other-wise $\bar{W}_{eq}(k)$ is non universal
implying that statistical mechanics does not exist in this case.
Inserting Eqs. (\ref{eq18},\ref{eq19}) in  Eq. (\ref{eq17})
we obtain
\begin{equation}
  \ln\left[\bar{W}_{eq}\left( k\right)\right] = - { T_2 k^2 \over 2 M} +
\left( { T_2 \over M } \right)^2 { q_4 - 3 \over 4! } 2 \epsilon k^4 + O\left( k^6 \right).
\label{eq19aaa}
\end{equation}
It is important to see that the $k^4$ term approaches
zero when $\epsilon \to 0$,
and hence
\begin{equation}
\lim_{\epsilon \to 0} \ln \left[ \bar{W}_{eq}(k) \right] = - {T_2 k^2 \over 2 M}.
\label{eq20}
\end{equation}
Thus the celebrated Maxwell--Boltzmann velocity PDF
is found
\begin{equation}
\lim_{\epsilon \to 0} W_{eq} \left( V_M  \right) = { \sqrt{M} \over \sqrt{ 2 \pi T_2} } \exp\left( - { M  V_M ^2 \over 2 T_2} \right).
\label{eq21}
\end{equation}
 Our  model exhibits the relation between
the central limit theorem and standard thermal equilibrium in
a very direct and simple way.
The main conditions
for thermal equilibrium are that
(i) number of collisions needed to reach equilibrium is very 
large namely $\epsilon$ is small, 
(ii) collisions are independent, (iii) scaling conditions
on velocity moments of bath particles must be satisfied, and
(iv) finite variance of velocity of bath particles PDF. 
In the next subsection we investigate possible power law generalizations of
standard equilibrium.

\begin{figure}[htb]
\epsfxsize=20\baselineskip
\centerline{\vbox{
        \epsfig{file=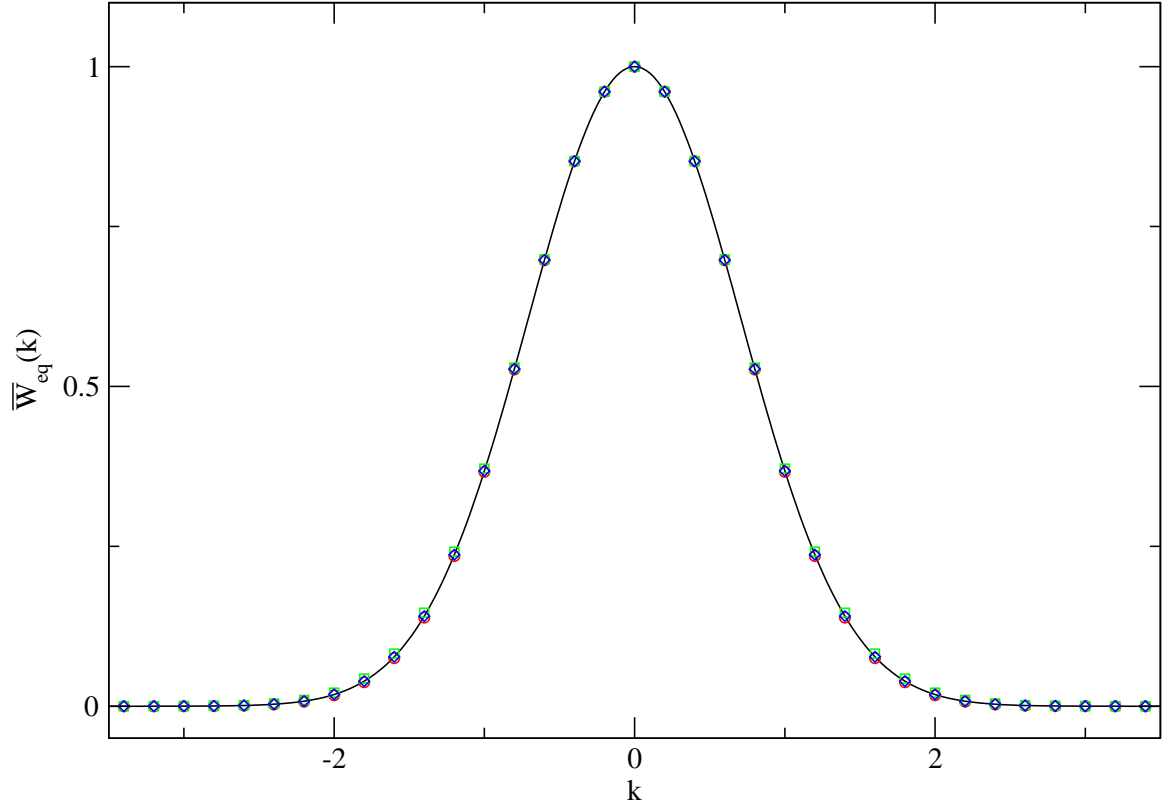, width=0.85\linewidth, angle=-90}  }}
\caption {
The  equilibrium characteristic function of the tracer particle,
$\bar{W}_{eq}(k)$ versus $k$.
We consider three types of bath particles
velocity PDFs
(i) exponential (squares),
(ii) uniform (circles), 
and (iii) Gaussian (diamonds).
The velocity distribution of the
tracer particle $M$ is well approximated by 
Maxwell's distribution plotted as the solid curve
 $\bar{W}_{eq}(k)= \exp\left( - |k|^{2}\right)$.
For the numerical results I used: $M=1$, $T_{2}=2$, $n=2000$,
 and $\epsilon=0.01$.
}
\label{fig1}
\end{figure}

{\bf remark 1} To fully characterize the equilibrium characteristic
function $\bar{W}_{eq}(k)$, the cumulant expansion
is investigated to infinite order. 
Let $\tilde{\kappa}_{m,2 j}$ $(\kappa_{M,2j})$ be the
$2 j$ th cumulant of bath particle (tracer particle) velocity,
 e.g. $\tilde{\kappa}_{m,2} = \langle \tilde{v}_m ^2 \rangle$,
$\tilde{\kappa}_{m,4} = \langle \tilde{v}_m ^4 \rangle
- \langle \tilde{v}_m ^2 \rangle^2$, etc. Then using Eq.
(\ref{eq13}) one can show that
\begin{equation}
\kappa_{M,2j} = g_{2 j} ^\infty \left( \epsilon \right) \tilde{\kappa}_{m, 2 j}.
\label{eq22}
\end{equation}
Since $g_{2 j} ^\infty(1) =1$ we find 
$\kappa_{M,2j} = \tilde{\kappa}_{m,2j}$  
if $m=M$, and hence for that case $\bar{W}_{eq}(k) = \bar{f}(k)$  as
mentioned previously.
 We assume that scaling properties of bath particles follow
$ \tilde{\kappa}_{m, 2 j}= c_{2 j} T_2 ^j/ m^j $, where $c_{2 j}$
are dimensionless parameters which depend on $f( \tilde{v}_m )$, 
$j=1,2,\cdots$,
and $c_{2 }=1$. The parameters $c_{2 j}$ for $j>1$ are 
the irrelevant parameters of the model in the limit of weak
collisions. To see this note that when $\epsilon \to 0$ we have 
$\kappa_{M,2 j} = (T_2/M) \delta_{j 1}$ implying relaxation
to a unique thermal equilibrium which is independent of the choice
of bath particle velocity distribution.

{\bf Remark 2} Instead of the condition Eq. (\ref{eq19})
it is sufficient to demand that 
the forth moment of bath particle velocity
will scale with the mass of the bath particles like
$ \langle \tilde{v}_m ^4 \rangle \propto {  1 \over m^\theta} $
and $ 0< \theta<3$.
If $\langle \tilde{v}_m ^4 \rangle \propto {  1 \over m^3} $
for example, we find that the second term in Eq.
(\ref{eq17}) does not vanish in the limit $\epsilon \to 0$.
In this case the equilibrium of the tracer particle will be
sensitive to behavior of $f(\tilde{v}_m )$,
and hence will exhibit a non-universal behavior.
\begin{figure}[htb]
\epsfxsize=20\baselineskip
\centerline{\vbox{
        \epsfig{file=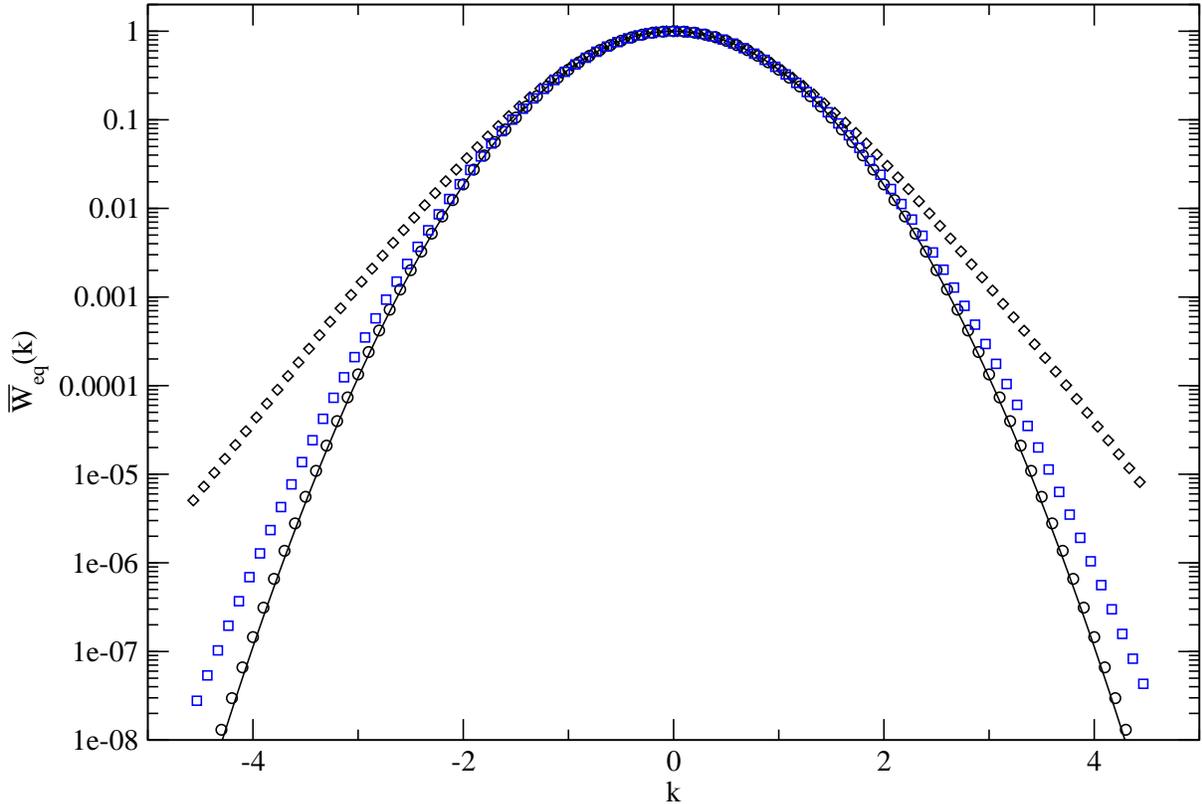, width=0.85\linewidth, angle=-90}  }}
\caption {
The tracer particle's equilibrium characteristic function,
for
the case when the velocity of the bath particles is exponentially
distributed Eq. 
(\protect{\ref{EqNE01}}).
The parameters are the same as in Fig. \protect{\ref{fig1}}
however now $\epsilon$ is varied. 
I use $\epsilon=0.05$ (diamonds),
$\epsilon=0.01$ (squares), $\epsilon=0.001$ (circles). As $\epsilon$
approaches zero, number of collisions needed to reach an equilibrium
becomes very large and then Maxwell--Boltzmann thermal equilibrium is obtained,
the solid curve $\bar{W}_{eq}(k)= \exp(- k^2)$. Note the logarithmic scale
of the figure.
}
\label{fig2}
\end{figure}

\subsection{L\'evy type of Equilibrium}

We now assume that the bath particle velocity PDF is even
with zero mean, and that it
decays like a power law
$P(\tilde{v}_m ) \propto |\tilde{v}_m| ^{-(1 + \alpha)}$
when  $\tilde{v}_m \to \infty$ where
$0<\alpha < 2$.
In this case the variance of
velocity distribution diverges. 
We use the small $k$ expansion
of the bath particle characteristic function  
\begin{equation}
\bar{f}\left( k \right) = 1 - {A_{\alpha} |k|^{\alpha} \over \Gamma\left( 1 + \alpha \right)} + { B |k|^{\beta} \over \Gamma\left( 1 + \beta \right) } +
o\left(|k|^{\beta} \right)
\label{eqL01}
\end{equation}
with $\alpha< \beta \le 2 \alpha$. 
Using Eq. (\ref{eq15}) we find
\begin{widetext}
\begin{equation}
 \ln \left[ \bar{W}_{eq}\left( k \right) \right] =  
\left\{
\begin{array}{l l}
-{ A_{\alpha}
\over \Gamma\left( 1 + \alpha \right)} 
 g_{\alpha} ^\infty \left( \epsilon \right) |k|^{\alpha} 
+ { B 
\over \Gamma\left( 1 + \beta \right) } 
g_{\beta} ^\infty \left(  \epsilon \right)|k|^\beta 
\ & \beta < 2 \alpha  \\
\ & \ \mbox{} \\
 -{A_{\alpha} 
 \over \Gamma\left( 1 + \alpha  \right)}
g_{\alpha} ^\infty \left( \epsilon \right)|k|^{\alpha} 
 + \left[ {B \over \Gamma\left( 1 + 2 \alpha \right)} - { A^2 \over 2 \Gamma^2 \left( 1 + \alpha\right) } \right] g_{2 \alpha} ^\infty \left( \epsilon \right) |k|^{2 \alpha} \ & \beta=2 \alpha \  \\
\end{array}
\right.
\label{eqL2}
\end{equation}
\end{widetext}
where terms of order higher than $|k|^{ \beta}$ are neglected.
Using Eq. (\ref{eq12a})
we obtain in the limit $\epsilon \to 0$
\begin{widetext}
\begin{equation}
 \ln \left[ \bar{W}_{eq} \left( k \right) \right] \sim 
\left\{
\begin{array}{l l}
- { A_{\alpha} \over \Gamma\left( 1 + \alpha \right) } \left( {m\over M} \right)^{\alpha -1} { 2^{\alpha -1} \over \alpha} |k|^{\alpha}
+ {B \over \Gamma\left( 1 + \beta \right)} \left( {m \over M} \right)^{\beta-1} {2^{\beta - 1} \over \beta } |k|^{\beta} & \beta  < 2 \alpha \\
\ & \ \\
-{A_{\alpha} \over \Gamma\left( 1 + \alpha \right) } \left( {m \over M} \right)^{\alpha -1} {2^{\alpha -1} \over \alpha} |k|^{\alpha} + 
\left[ {B \over \Gamma\left( 1 + 2 \alpha \right) } - {A^2  \over 2 \Gamma^2\left( 1 + \alpha \right) } \right] \left( {m \over M} \right)^{2 \alpha -1} {2^{ 2 \alpha -1} \over 2 \alpha} |k|^{2 \alpha} & \beta= 2 \alpha.
\end{array}
\right.
\label{eqL03}
\end{equation}
\end{widetext}
To obtain an equilibrium we use the thermodynamical argument.
We require that equilibrium velocity PDF of the tracer
particle $M$ be independent of the mass of the bath particle,
as we did for the Maxwell--Boltzmann case
Eq. (\ref{eq18}).
This
condition yields
\begin{equation}
A_{\alpha} = {T_{\alpha} \over m^{\alpha -1} },
\label{eqL04}
\end{equation}
where within the context of our model $T_{\alpha}$
has a meaning of generalized temperature whose units are
$\mbox{Kg}^{\alpha -1} \mbox{Mt}^{\alpha} / \mbox{Sec}^{\alpha} $. 
An additional requirement is needed to obtain a unique equilibrium
(i.e., an equilibrium which does not depend on
details of bath particle velocity PDF like $B$):
terms beyond the $|k|^{\alpha}$ term in Eq.
(\ref{eqL03})
must  vanish in the limit
$\epsilon \to 0$. This occurs for a large family of velocity 
PDFs $f\left( \tilde{v}_m \right)$ which satisfy the condition
\begin{equation}
B \propto {1 \over m^{\theta} }
\label{eqL05}
\end{equation}
and now $\theta < \beta - 1$. For the Gauss-Maxwell-Boltzmann case
considered in previous section we had $\beta=4$  while now
$\beta<4$.
Using this condition we obtain 
\begin{equation}
\lim_{\epsilon \to 0}\  \ln \left[ \bar{W}_{eq} \left( k \right) \right] = 
- {T_{\alpha} 2^{\alpha -1}  \over \Gamma\left( 1 + \alpha \right) \alpha M^{\alpha-1}}|k|^{\alpha},
\label{eqL06}
\end{equation}
hence L\'evy type of equilibrium is obtained \\
\begin{equation}
  \lim_{\epsilon \to 0} W_{eq} \left( V_M \right) = 
\left[ { \Gamma\left( 1 + \alpha \right) M^{*(\alpha-1)} \over T_{\alpha} }\right]^{1/\alpha} 
l_{\alpha} \left\{ \left[ { \Gamma\left( 1 + \alpha \right) M^{*(\alpha -1)}  \over T_{\alpha} }\right]^{1/\alpha} V_M \right\}
\label{eqL07}
\end{equation}
where $M^*= M \alpha^{1/(\alpha -1)}/2$ is the renormalized mass.
For $\alpha =2$ the Maxwell--Boltzmann PDF Eq. 
(\ref{eq21}) is recovered and $M^* = M$.

 From physical requirements we note that
 our results are valid only when $1<\alpha\le 2$.
If $\alpha=1$ the velocity PDF
becomes independent of the mass of the tracer particle,
while when $\alpha< 1$ the heavier the tracer particle
the faster its motion (in statistical sense). This
implies that imposing the condition of independence
of tracer particle velocity PDF on the coupling constant, based
on the thermodynamic argument, is not the correct path.
This is one of the reasons why I consider a scaling
approach in the next section. The unphysical behavior
also indicates that the assumption of
uniform collision rate is unphysical 
when $\langle |\tilde{v}_m| \rangle = \infty$, implying
that the mean field approach we are using breaks down
when $\alpha\le 1$. 

{\bf Remark 1} The domain of attraction of the L\'evy equilibrium
we find, Eq. (\ref{eqL07}) does not include all power
law distribution with $\alpha<2$. 
Consider for example 
\begin{equation}
\bar{f}(k) ={1 \over 2} \left\{\exp\left[  - {2 T_{\alpha} |k|^{\alpha}\over m^{\alpha-1}\Gamma\left(1+\alpha\right)} \right] + \exp\left( - {T_2 k^2 \over  m }\right)\right\},
\label{eqeexx}
\end{equation}
hence the gas particle velocity distribution in this case is a sum of
a Gaussian and a L\'evy distributions. The small $k$
expansion of Eq. (\ref{eqeexx}) is 
\begin{equation}
\bar{f}(k) = 1 - { T_{\alpha} |k|^{\alpha}\over m^{\alpha-1}\Gamma\left(1+\alpha\right)} -  {T_2 k^2 \over 2  m } \cdots, 
\label{eqeexx1}
\end{equation}
where we consider $1 < \alpha < 2$. 
Using Eqs. (\ref{eqL01},\ref{eqeexx1}) we find
 $\beta=2$, while comparing
Eq. 
(\ref{eqL05}) and Eq. (\ref{eqeexx1}) yields $\theta=1$.
Now the condition $\theta < \beta - 1$ does not hold, and 
hence the L\'evy equilibrium 
Eq. (\ref{eqL07}) is not obtained.
Interestingly, one can show that for this case one obtains
an equilibrium characteristic function
$\bar{W}_{eq}(k)$  which is a convolution of a L\'evy PDF
and a  Gaussian PDF. I suspect that this type of equilibrium
is not limited to this example.

\subsection{Numerical Examples}

 To demonstrate the results
exact solutions of the problem are investigated.
This yields insight into convergence rate
to stable equilibrium. 

\subsubsection{Maxwell Statistics}

 First consider the Maxwell--Boltzmann case. We investigate three
types of bath particle velocity PDFs:\\
$(i)$ The exponential
\begin{equation}
f\left( \tilde{v}_m \right) = {\sqrt{2 m} \over 2 \sqrt{T_2} } \exp\left( -
{\sqrt{2 m} |\tilde{v}_| \over \sqrt{T_2} } \right),
\label{EqNE01}
\end{equation}
which yields 
\begin{equation}
\bar{f}(k) = {1 \over 1 + { T_2 k^2 \over 2 m} }.
\label{EqNE02}
\end{equation}
$(ii)$ The uniform PDF
\begin{equation}
f(\tilde{v}_m) =
\left\{
\begin{array}{l l}
 \sqrt{ {m \over 12 T_2 } } \ & \mbox{if} \ |\tilde{v}_m|< \sqrt{{ 3 T_2 \over m}}\\ 
\ & \ \\
 0 \ & \ \mbox{otherwise} \\ 
\end{array}
\right.
\label{EqNE03}
\end{equation}
which yields
\begin{equation}
\bar{f}(k) = { \sin \left( \sqrt{ {3 T_2 \over m}}  k \right) \over 
\sqrt{ { 3 T_2 \over m}} k } .
\label{EqNE04}
\end{equation}
$(iii)$ The Gaussian PDF
\begin{equation}
\bar{f}(k) = \exp\left( - {k^2 T_2 \over 2 m } \right).
\label{EqNE05}
\end{equation}
 The small $k$ expansion of Eqs. (\ref{EqNE02},\ref{EqNE04},\ref{EqNE05})
is $\bar{f}(k) \sim 1 - k^2 T_2 / ( 2 m) + \cdots$, indicating that the
second moment of velocity of bath particles $\langle \tilde{v}_m ^2 \rangle$
is identical for the
three PDFs. 

 To obtain numerically exact solution of the problem we use Eq. 
(\ref{eq13}) with large though finite $n$. In all our numerical examples
we used
$M=1$ hence $m=\epsilon$. Thus for example for
the uniform velocity PDF Eq. 
(\ref{EqNE04})
  we have
\begin{equation}
 \bar{W}_{eq} \left( k \right) \simeq 
\exp \left\{ \sum_{i=1}^n \ln 
\left[ \sqrt{ {\epsilon \over 3 T_2}} 
{ \sin \left( k \sqrt{{ 3 T_2 \over m}} \left( { 1 - \epsilon \over 1 + \epsilon } \right)^{n-i} { 2 \epsilon \over 1 + \epsilon} \right)\over 
k \left( { 1 - \epsilon \over 1 + \epsilon } \right)^{n-i} { 2 \epsilon \over 1 + \epsilon } }
\right] \right\}.
\label{EqNE06}
\end{equation}
To obtain equilibrium we
increase $n$ for a fixed $\epsilon$ and temperature until
a stationary solution is obtained. 

 According to our analytical results the 
bath particle velocity PDFs Eqs. (\ref{EqNE01},\ref{EqNE03},\ref{EqNE05}),
 belong to the domain of attraction of the Maxwell--Boltzmann
equilibrium.
 In Fig.
\ref{fig1} we show $\bar{W}_{eq}(k)$ obtained from numerical
solution of the problem. The numerical solution 
exhibits an excellent agreement
  with the 
Maxwell--Boltzmann equilibrium. 
Thus details of the precise shape of
velocity PDF of bath particles are unimportant,
and as expected the Maxwell-Boltzmann distribution is stable.
To obtain the results in Fig. \ref{fig1}
I used $\epsilon=0.01$, $T_2 = 2$, $M=1$ and $n=2000$.
For convenience  Fourier space is used, 
 the solid curve is  theoretical prediction
$ \ln[W_{eq} (k)]=  - k^2$ which is strictly valid in the limit 
$n \to \infty$ and then $\epsilon \to 0$.

A closer look at $\bar{W}_{eq}( k ) $ reveals that
for large $k$ deviations from Gaussian behavior
are found for finite values of $\epsilon$, as might
be expected.
In Fig. \ref{fig2} we show  the numerically exact solution
using the  exponential velocity PDF Eq. (\ref{EqNE01})
and vary the mass ratio $\epsilon$.
The figure demonstrates that when $\epsilon \to 0$ 
the asymptotic theory becomes exact, the convergence rate depends on the
value of $k$.
 As we noted already the Maxwell--Boltzmann statistics is reached
only if number of collisions needed to obtain an equilibrium
is very large, which implies small values of
$\epsilon$.  

\begin{figure}[htb]
\epsfxsize=20\baselineskip
\centerline{\vbox{
        \epsfig{file=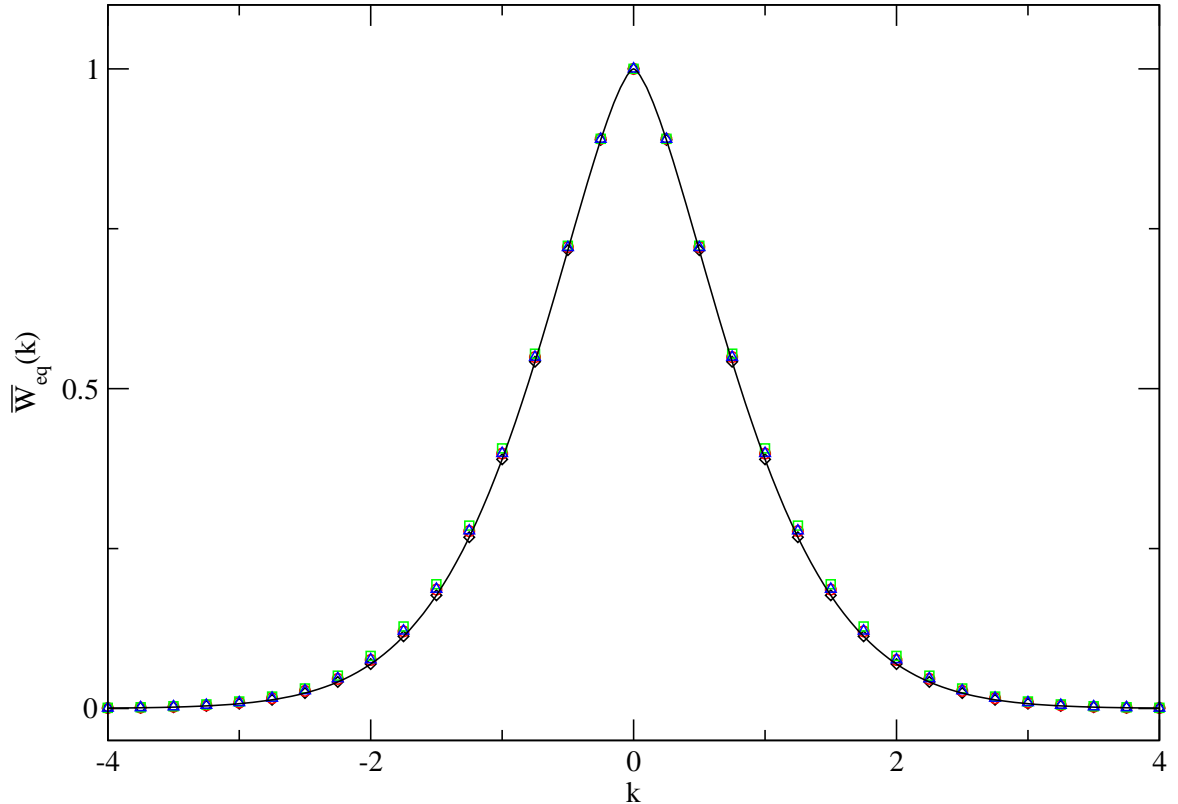, width=0.85\linewidth, angle=-90}  }}
\caption {
The equilibrium characteristic function $\bar{W}_{eq}(k)$,
for the case when velocities of bath particles 
are distributed according to a power law with $\alpha=3/2$.
Four types of bath particles
characteristic functions are considered:
(i)
the Hyper-geometric function Eq. \protect{(\ref{EqNE07})}
 (squares),
(ii) 
Meijer G function Eq. \protect{(\ref{EqNE08})},
(circles), (iii)
Bessel function Eq. \protect{(\ref{eqTs01})},
(triangles),
and (iv) the Holtsmark function  Eq.
\protect{(\ref{EqNE09})}
(diamonds).
For all these cases, the equilibrium characteristic function $\bar{W}_{eq}(k)$
is well approximated by the L\'evy characteristic function;
the solid curve
 $\bar{W}_{eq}(k)= \exp\left( - 2^{3/2} |k|^{3/2}/3\right)$.
For the numerical results we used: 
$M=1$, $T_{\alpha}=\Gamma(5/2)$, $\epsilon=5e-5$, and $n=1e6$. 
}
\label{fig3}
\end{figure}

\subsubsection{L\'evy Statistics}

 We now consider four power law PDFs satisfying
$f\left( \tilde{v}_m \right) \propto |\tilde{v}_m |^{ - 5/2 }$,
namely $\alpha=3/2$. \\
$(i)$ Case $1$ we choose
\begin{equation}
f\left( \tilde{v}_m \right)={ N_1 \over \left( 1 + c_1 |\tilde{v}_m|\right)^{5/2}},
\label{EqNE07}
\end{equation}
with $c_1= \left( 3 \pi \right)^{2/3} m^{1/3} 2^{-1} (T_{3/2})^{2/3} $
and $N_1= 3 c_1/4$. The characteristic function is given
in terms of a Hyper-geometric function \cite{Math}
%
\begin{widetext}
\begin{equation}
 \bar{f}(k)= 
    \frac{4 N_1\,\left(  c_1 ^{\frac{3}{2}}\,
            \mbox{HypergeometricPFQ}(\{ 1\} ,
             \{ -\left( \frac{1}{4} \right) ,\frac{1}{4}\} ,
             \frac{-k^2}{4\,c_1 ^2})   - 
         {| k|}^{\frac{3}{2}}\,{\sqrt{2\,\pi }}\,
          \left( \cos (\frac{k}{c_1}) + 
            \mbox{Sign}(k)\,\sin (\frac{k}{c_1}) \right)  \right) }{3\,
       c_1 ^{\frac{5}{2}}}.
\label{EqMath1}
\end{equation}
\end{widetext}
$(ii)$ Case $2$ 
\begin{equation}
f\left( \tilde{v}_m \right) = { N_2 \over  1 + c_2 |\tilde{v}_m|^{5/2} } ,
\label{EqNE08}
\end{equation}
where $c_2 = ( 2 \sqrt{ 2 \pi m} )^{5/3} ( 3 G T_{3/2} )^{-5/3} $,
$G=(4 \pi/5) \sqrt{ 2/ ( 5 + \sqrt{5} ) }$, $N_2=c_2 ^{2/5}/ (2 G)$.
The characteristic function is expressed in terms
of Meijer G function  \cite{Math}
\begin{widetext}
\begin{equation}
 \bar{f} ( k ) = 
   \frac{ 2 N_2 \mbox{MeijerG}(\{ \{ \frac{3}{20},\frac{2}{5},\frac{13}{20}\} ,
       \{ \} \} ,\{ \{ 0,\frac{3}{20},\frac{1}{5},\frac{2}{5},\frac{2}{5},
        \frac{3}{5},\frac{13}{20},\frac{4}{5},\frac{9}{10}\} ,
       \{ \frac{1}{10},\frac{3}{10},\frac{1}{2},\frac{7}{10}\} \} ,
      \frac{k^{10}}{10000000000\,c_2 ^4})}{4\,{\sqrt{5}}\,c_2 ^{\frac{2}{5}}\,
      {\pi }^{\frac{5}{2}}}.
\label{EqMath2}
\end{equation}
\end{widetext}
To obtain $\bar{W}_{eq}(k)$ we used the numerical
Fourier transform of Eqs. 
(\ref{EqNE07},\ref{EqNE08}), instead of the rather formal exact expressions 
Eqs. (\ref{EqMath1},\ref{EqMath2}). \\
(iii)  Case $3$
\begin{equation}
f\left( \tilde{v}_m \right) = { N_3 \over \left( 1 + C_3 \tilde{v}_m ^2\right)^{5/4}}
\label{eqTs01}
\end{equation}
where 
$C_3=\{\Gamma[1/4]\Gamma[5/2] \sqrt{2 m} 3^{-1} 
T_{3/2} ^{-1} \Gamma^{-1} (3/4)]\}^{4/3}$,
$N_3= 0.75 T_{3/2} \Gamma^{-1}(5/2) ( 2 \pi m)^{-1/2} C_3 ^{5/4}$. 
The characteristic function is
\begin{equation}
\bar{f}(k) = N_3 C_3 ^{-7/8} {2^{1/4}\sqrt{\pi} \over \Gamma\left( { 5 \over 4} \right) } |k|^{3/4} K_{3/4} \left( { | k | \over \sqrt{C_3} } \right), 
\label{eqTs02}
\end{equation}
where $K_{3/4}$ is the modified Bessel function of the 
second kind. \\
$(iv)$ Case $4$ the L\'evy PDF with index $3/2$ whose
Fourier pair is
\begin{equation}
\bar{f}\left( k \right) = \exp \left[ - { T_{3/2} |k|^{5/2} \over \sqrt{m} \Gamma\left( 5/2 \right)} \right].
\label{EqNE09}
\end{equation}

The $|\tilde{v}_m| \to \infty$ behavior of the PDFs
(\ref{EqNE07}-\ref{EqNE09}) is
\begin{equation}
f\left( \tilde{v}_m \right) \sim { 3 T_{3/2} \over 4 \sqrt{2 \pi m} } | \tilde{v}_m |^{-5/2} ,
\label{EqNE10}
\end{equation}
hence the small $k$ behavior of the characteristic function is
\begin{equation}
\bar{f}\left( k \right) \sim 1 - { T_{3/2} | k|^{3/2} \over \sqrt{m} \Gamma\left( 5/2 \right) } .
\label{EqNE11}
\end{equation}
According to our results in previous section the velocity PDFs
Eqs. (\ref{EqNE07}-\ref{EqNE08}) belong to the domain of attraction
of the L\'evy type of equilibrium 
\begin{equation}
\lim_{\epsilon \to 0} \bar{W}_{eq} \left( k \right) = \exp \left[ - { T_{3/2} 2^{3/2} \over \sqrt{M} 3 \Gamma\left( 5/2 \right) } | k|^{3/2} \right].
\label{EqNE12}
\end{equation}
Namely, $W_{eq}\left( V_M \right)$ is the L\'evy PDF with
index $3/2$ also called the Holtsmark PDF. 

\begin{figure}[htb]
\epsfxsize=20\baselineskip
\centerline{\vbox{
        \epsfig{file=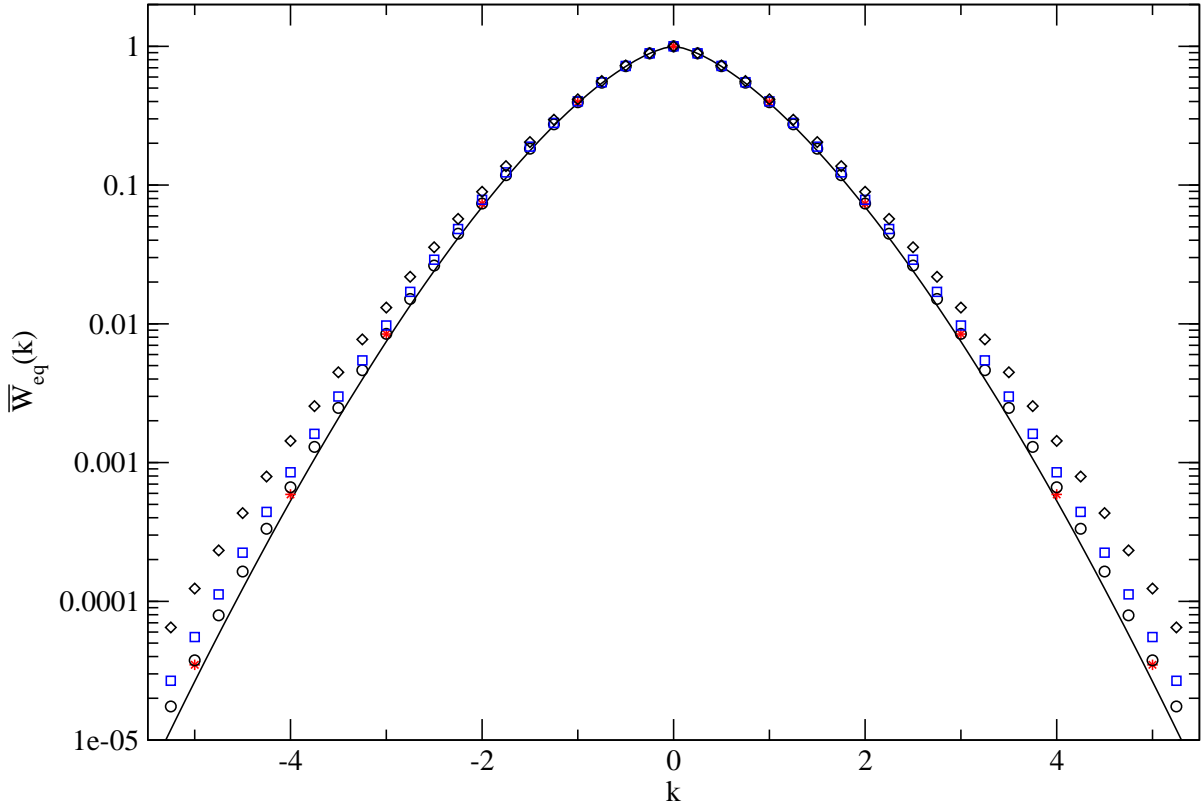, width=0.85\linewidth, angle=-90}  }}
\caption {
The tracer particle equilibrium characteristic function, for
the case when the characteristic function of the bath particles
is described
by the Bessel function  Eq. 
(\protect{\ref{eqTs02}}).
The parameters are the same as in Fig. \protect{\ref{fig3}}
however now $\epsilon$ is varied. 
We use $\epsilon=10^{-3}$ (diamonds),
$\epsilon=10^{-4}$ (squares), $\epsilon=10^{-5}$ (circles),
and $\epsilon=10^{-6}$ (stars). As $\epsilon$
approaches zero, number of collisions needed to reach an equilibrium
becomes very large and then a L\'evy type of equilibrium is obtained,
the solid curve 
 $\bar{W}_{eq}(k)= \exp\left( - 2^{3/2} |k|^{3/2}/3\right)$.
 Note the logarithmic scale
of the figure.
}
\label{fig4}
\end{figure}

 Similar to the Maxwell-Boltzmann case,  numerically exact 
solution are obtained, in $k$ space
using Eq. (\ref{eq13}) with finite $n$. For example for 
the power law velocity PDF Eq. (\ref{eqTs01}) we use
 $$ \bar{W}_{eq} \left( k \right) \simeq  $$
\begin{widetext}
\begin{equation}
\exp\left\{ \sum_{i=1} ^n \log\left[ N_3 C_3 ^{-7/8} { 2^{1/4} \sqrt{\pi} \over \Gamma\left( 5/4 \right) } |k|^{3/4} \left( { 1 - \epsilon \over 1 + \epsilon } \right)^{3(n-i)/4} \left( { 2 \epsilon \over 1 + \epsilon } \right)^{3/4} K_{3/4} \left( \left( { 1 - \epsilon \over 1 + \epsilon}\right)^{n-i} |k| { 2 \epsilon \over 1 + \epsilon} \left(C_3\right)^{-1/2} \right) \right] \right\}, 
\label{EqNE09f}
\end{equation}
\end{widetext}
where we fix $\epsilon$ and $T_{3/2}$
 and increase $n$ until a stationary solution
is obtained.

 In Fig. \ref{fig3} we show $\bar{W}_{eq}(k)$ obtained using numerically
exact solution based on the four PDFs Eqs. 
(\ref{EqNE07},\ref{EqNE08},\ref{eqTs01},\ref{EqNE09}).
The exact solutions are in good agreement with
the theoretical prediction 
Eq. (\ref{EqNE12}). Thus, similar to the Maxwell--Boltzmann case, the
exact shape of the equilibrium distribution does not 
depend on the details of the velocity PDF of the bath particles (besides
$\alpha$ and $T_{\alpha}$ of-course).

A closer look at $\bar{W}_{eq}( k ) $ reveals that
for large $k$ deviations from L\'evy behavior
are found for finite values of $\epsilon$, similar to what 
we have shown for the Gaussian case Fig. \ref{fig2}.
In Fig. \ref{fig4} we show  the numerically exact solution
using the power law velocity PDF  Eq. 
(\ref{eqTs01})
and vary the mass ratio $\epsilon$.
The figure demonstrates that when $\epsilon \to 0$ 
the asymptotic theory becomes exact, the convergence rate depends on the
value of $k$.
 The L\'evy type of equilibrium is reached
only if number of collisions needed to obtain an equilibrium
is very large, which implies small values of
$\epsilon$.  
The convergence towards a stable equilibrium was found to be slow
compared with the Gaussian case
(for $\epsilon = 1e-6$ we used
$n=3e6$ to obtain a stationary solution).

\begin{figure}[htb]
\epsfxsize=20\baselineskip
\centerline{\vbox{
        \epsfig{file=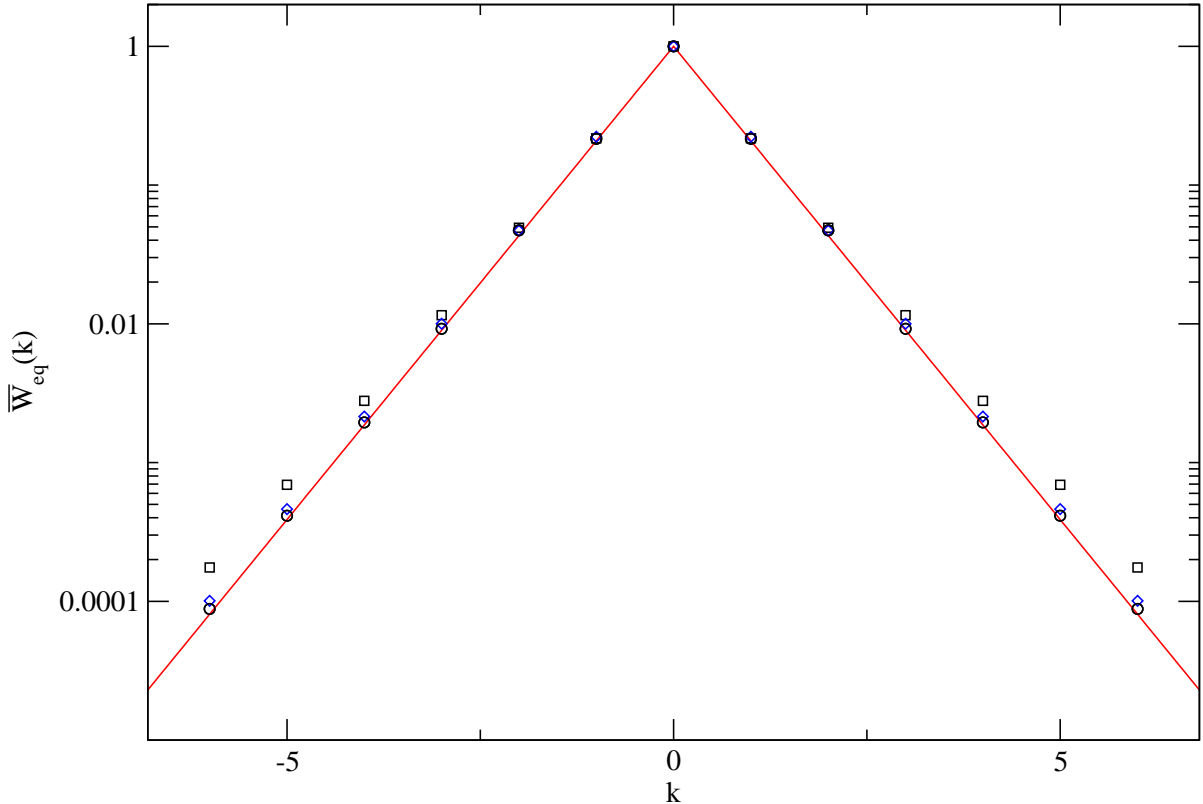, width=0.85\linewidth, angle=-90}  }}
\caption {
The equilibrium characteristic function, for
a case when the velocity of the bath particles 
is a power law 
with $\alpha=1$ Eq.
(\ref{eq52}). 
We use $\epsilon=0.01$ (squares),
$\epsilon=0.001$ (diamonds), $\epsilon=0.0005$ (circles). As $\epsilon$
approaches zero, number of collisions needed to reach an equilibrium
becomes very large and then a L\'evy type of
 thermal equilibrium is obtained:
the solid curve 
$\bar{W}_{eq}(k)=\exp(- \pi |k|/2)$. Note the logarithmic scale
of the figure.
}
\label{fig5}
\end{figure}

 We also consider the marginal case $\alpha=1$, which marks
the transition from finite $\langle | \tilde{v}_m | \rangle$ for $\alpha>1$
to infinite value of $\langle | \tilde{v}_m | \rangle$ for $\alpha<1$.
We considered the velocity PDF
\begin{equation}
f\left( \tilde{v}_m \right) = {1 \over 2 
\left( 1 + |\tilde{v}_m|\right)^{2} } .
\label{eq52}
\end{equation}
The  characteristic function  is \cite{Math}
\begin{equation}
\bar{f} \left( k \right) =
   \frac{\mbox{MeijerG}(\{ \{ 0\} ,\{ \} \} ,
      \{ \{ 0,\frac{1}{2},1\} ,\{ \} \} ,\frac{k^2}{4})}{{\sqrt{\pi }}},
\label{eqxx01}
\end{equation}
which for small $k$ behaves like 
$\bar{f} \left( k \right) = 1 - \pi |k|/2\cdots$. 
According to the theory in the limit $\epsilon \to 0$
\begin{equation}
\bar{W}_{eq}(k) = \exp( - \pi |k|/2).
\label{eqxx02}
\end{equation}
Thus the velocity PDF of the tracer 
particle $M$ is the Lorentzian
density. 
As mentioned in this case the equilibrium obtained is independent of
mass $M$.
In Fig.
\ref{fig5} we show the numerically exact solution of 
$\bar{W}_{eq}\left( k \right)$ for several mass ratios 
$\epsilon$. As $\epsilon \to 0$ we obtain
the predicted L\'evy type of equilibrium.

\section{Scaling Approach: the Relevant Scale is Energy}
\label{SecSca}

A second approach to the problem is now considered.
We assume that statistical properties of bath particles
velocities can be characterized with an energy scale $T$.
Since $T$, $m$ and $\tilde{v}_m$ are the only variables
describing the bath particle we have
\begin{equation}
f\left( \tilde{v}_m \right) = {1 \over \sqrt{ T / m} } q\left( {\tilde{v}_m \over \sqrt{ T/ m } } \right).
\label{eqSca01}
\end{equation}
We also assume that $f(\tilde{v}_m )$ is an even function,
as expected from symmetry. The dimensionless function
$q(x)\ge 0$ satisfies a normalization condition
\begin{equation}
\int_{-\infty}^{\infty} q(x) {\rm d} x = 1,
\label{eqSca02}
\end{equation}
otherwise it is rather general.
The scaling assumption made in Eq. (\ref{eqSca01}) is very natural,
since the total energy of bath particles is nearly conserved,
i.e. the energy transfer to the single heavy particle being
much smaller than the total energy of the bath particles.
From this starting point we investigate now equilibrium properties of
the tracer particle $M$. Showing among
other things that $T$ has the meaning of temperature.

\subsection{ Gauss--Maxwell--Boltzmann Distribution}

 We first consider the case where moments of $f\left( \tilde{v}_m \right)$
are finite. The second moment of the bath particle velocity is
\begin{equation}
\langle \tilde{v}_m ^2 \rangle = { T \over m} \int_{-\infty}^{\infty} x^2 q(x) {\rm d} x.
\label{eqSca03}
\end{equation}
Without loss of generality we set 
$\int_{-\infty}^{\infty} x^2 q(x) {\rm d} x=1$.
The scaling behavior Eq. 
(\ref{eqSca01}) and the assumption of finiteness of moments
of the PDF yields
\begin{equation}
\langle \tilde{v}_m ^{2 n} \rangle = \left( { T \over m} \right)^n q_{2 n},
\label{eqSca04}
\end{equation}
where the moments of $q(x)$ are defined according to
\begin{equation}
q_{2n} = \int_{-\infty}^{\infty} x^{2 n} q(x) {\rm d} x.
\label{eqSca05}
\end{equation}
Thus the small $k$ expansion of the characteristic function is
\begin{equation}
 \bar{f}\left( k \right) = 
 1 - { T k^2 \over 2 m } + q_4 \left( { T \over m} \right)^2 { k^4 \over 4!} + O(k^6).
\label{eqSca06}
\end{equation}
Inserting Eq. (\ref{eqSca06}) in Eq.  
(\ref{eq13})
we obtain
\begin{equation}
 \ln \left[ \bar{W}_{eq}\left(k\right) \right] = 
- { T \over 2 m} g_2 ^{\infty} \left( \epsilon \right) k^2 +
{q_4 -3 \over 4!} \left( {T \over m} \right)^2 g_4 ^\infty\left( \epsilon \right) k^4 + O(k^6).
\label{eqSca07}
\end{equation}
In the limit $\epsilon \to 0$ the second term on the right hand
side of Eq. (\ref{eqSca07}) is zero, using Eq. 
(\ref{eq12b})
\begin{equation}
 \lim_{\epsilon \to 0} \ln \left[ \bar{W}_{eq}\left(k\right) \right] = 
- { T \over 2 M} k^2. 
\label{eqSca07a}
\end{equation}
The parameters $q_{2 n}$ with $n> 1$ are the irrelevant parameters 
of the problem. From Eq. (\ref{eqSca07a}) it is easy
to see that  the Maxwell--Boltzmann velocity PDF for the tracer particle
$M$ is obtained. 

\subsection{L\'evy Velocity Distribution}

 Now we assume that the variance of  $f\left( \tilde{v}_m \right)$
diverges, i.e., $q(x)\propto |x|^{-(1 + \alpha)}$ when $|x| \to \infty$
and $0<\alpha<2$. For this case the bath particle characteristic function
is
\begin{equation}
 \bar{f}\left( k \right) = 
1 - { q_{\alpha} \over \Gamma\left( 1 + \alpha \right) } \left( { T \over m} \right)^{\alpha/2} |k|^{\alpha} 
+ { q_{\beta} \over \Gamma\left( 1 + \beta \right) } \left( { T \over m} \right)^{\beta/2} |k|^{\beta} 
+ o(|k|^{\beta})
\label{eqSca07b}
\end{equation}
where $\alpha< \beta \le 2 \alpha$.
 In many cases $\beta = 2$, an example is given
in next subsection. $q_{\alpha}$ and $q_{\beta}$ are dimensionless numbers
which depend of-course on $q(x)$. Without loss of generality we may
set $q_{\alpha} = 1$.
In Eq. (\ref{eqSca07b}) we have used the assumption that $f(\tilde{v}_m)$
is even.

 Using the same technique used in previous
section we obtain the equilibrium characteristic 
function for the tracer particle $M$
\begin{widetext}
\begin{equation}
 \ln \left[ \bar{W}_{eq}\left( k \right) \right] =  
\left\{
\begin{array}{l l}
-{ 1
\over \Gamma\left( 1 + \alpha \right)} 
\left( { T \over m} \right)^{\alpha/2}  g_{\alpha} ^\infty \left( \epsilon \right) |k|^{\alpha} 
+ { q_{\beta} 
\over \Gamma\left( 1 + \beta \right) } 
\left( {T \over m} \right)^{\beta/2} g_{\beta} ^\infty \left(  \epsilon \right)|k|^\beta 
 & \beta < 2 \alpha  \\
\ & \ \mbox{} \\
 -{ 1
 \over \Gamma\left( 1 + \alpha  \right)}
\left( { T \over m} \right)^{\alpha/2} g_{\alpha} ^\infty \left( \epsilon \right)|k|^{\alpha} 
 + \left[ {q_{2 \alpha} \over \Gamma\left( 1 + 2 \alpha \right)} - { 1 \over 2 \Gamma^2 \left( 1 + \alpha\right) } \right] \left( { T \over m} \right)^{\alpha} g_{2 \alpha} ^\infty \left( \epsilon \right) |k|^{2 \alpha} \ & \beta=2 \alpha. \  \\
\end{array}
\right.
\label{eqSca08}
\end{equation}
\end{widetext}
Taking the small  $\epsilon$ limit the following expansions are found,
if $\beta< 2 \alpha$ 
\begin{widetext}
\begin{equation}
 \ln \left[ \bar{W}_{eq} \right( k \left) \right] \sim
- { 2^{\alpha -1} \over \alpha \Gamma\left( 1 + \alpha \right) } 
\left( { T \over M } \right)^{\alpha/2} { |k|^{\alpha} \over \epsilon^{1 - \alpha/2} } +
{ q_{\beta} 2^{\beta -1} \over 2 \Gamma\left( 1 + \beta \right) } \left( { T \over M} \right)^{\beta/2} { |k|^{\beta} \over \epsilon^{1 - \beta/2} } + o(|k|)^\beta,
\label{eqSca09}
\end{equation}
\end{widetext}
if $\beta=2 \alpha$ 
\begin{widetext}
\begin{equation}
\ln \left[ \bar{W}_{eq} \right( k \left) \right] \sim
- { 2^{\alpha -1} \over \alpha \Gamma\left( 1 + \alpha \right) } 
\left( { T \over M } \right)^{\alpha/2} { |k|^{\alpha} \over \epsilon^{1 - \alpha/2} } +
\left[ { q_{\beta}  \over  \Gamma\left( 1 + \beta \right) } - { 1 \over 2 \Gamma^2 \left( 1 + \alpha\right)} \right] \left( { T \over M} \right)^{\alpha}
{2^{ 2 \alpha -1} \over 2 \alpha} { |k|^{2 \alpha} \over \epsilon^{1 - \alpha} } + o(|k|^{2\alpha} ).
\label{eqSca10}
\end{equation}
\end{widetext}
Thus for example if $\beta=2$ the leading term in Eq. (\ref{eqSca09})
scales with $\epsilon$ according to $\epsilon^{\alpha/2 -1} \to \infty$,
while the second term scales like $\epsilon^0$.  Thus
from Eqs. (\ref{eqSca09},\ref{eqSca10}) we see
when $\epsilon$ is small and $k$ not too large, we may neglect
the second and higher order terms. This yields 
a L\'evy type of behavior
\begin{equation}
\bar{W}_{eq} \left( k \right) \sim \exp\left[ - { 2^{\alpha-1} \over \alpha \Gamma\left( 1 + \alpha\right) } \left( { T \over M} \right)^{\alpha/2} {|k|^{\alpha} \over \epsilon^{1 - \alpha/2} } \right]. 
\label{eqSca100}
\end{equation} 
We note
that for $\alpha \ne 2$
the equilibrium Eq. (\ref{eqSca100}) depends
on $\epsilon$. While for the Maxwell--Boltzmann case $\alpha=2$, 
the equilibrium is independent of the coupling constant $\epsilon$.
This difference between the L\'evy equilibrium and 
the Maxwell--Boltzmann equilibrium is related to
the conservation of energy and momentum during
a collision event, and to the fact that the
energy of the particles is quadratic in their velocities.
The asymptotic behavior Eq. (\ref{eqSca100}) is now 
demonstrated using  numerical examples.

\begin{figure}[htb]
\epsfxsize=20\baselineskip
\centerline{\vbox{
        \epsfig{file=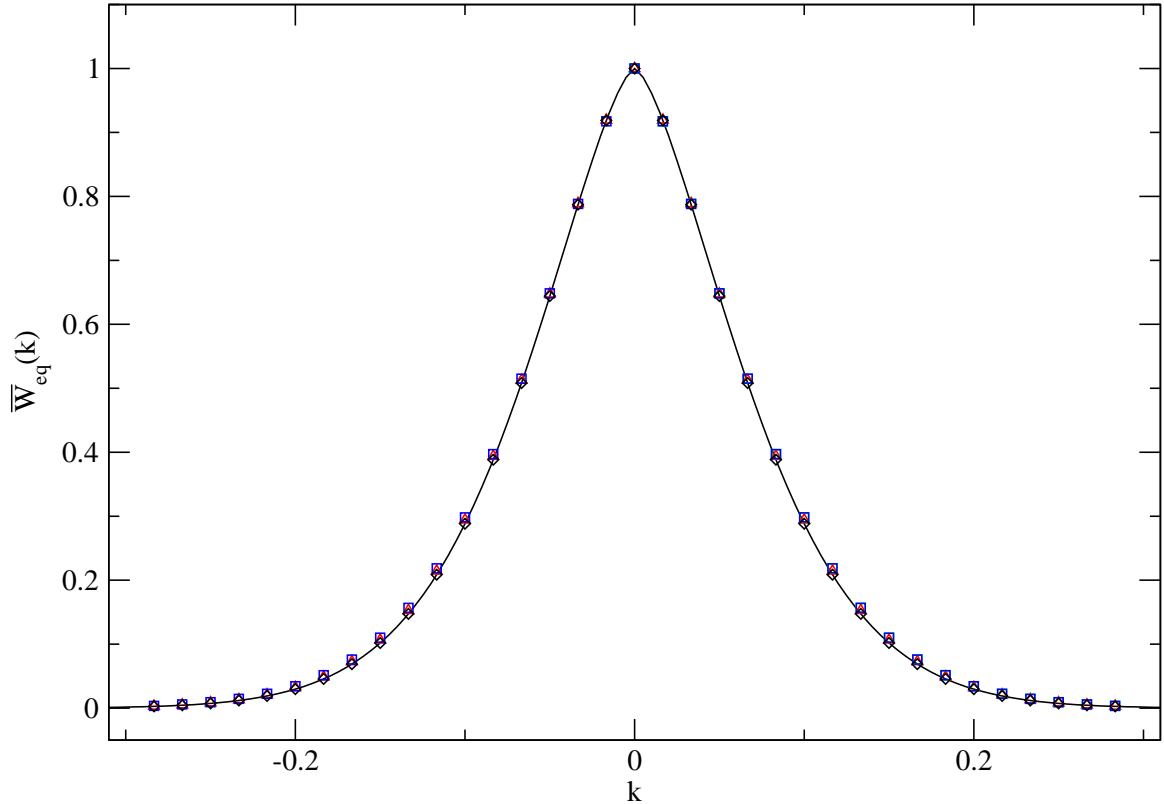, width=0.85\linewidth, angle=-90}  }}
\caption {
We show the equilibrium characteristic function
of the tracer particle, 
using the scaling approach.
Numerically exact solution
of the problem are obtained using three long tailed 
bath particle
velocity PDFs  
(i) Eq.  
(\protect{\ref{eqexs1}})
squares, (ii) Eq. 
(\protect{\ref{eqEN01}})
triangles, (iii) Eq.  
(\protect{\ref{eqLLVV}}) diamonds.
The tracer particle equilibrium
is well approximated by the L\'evy characteristic function
the solid curve;
$\bar{W}_{eq}(k)\sim\exp\left(-2.211|{k\over \epsilon^{1/6} }|^{3/2} \right)$. 
For the numerical results we used $T=4.555$, $\epsilon=1e-5$ and $M=1$.
}
\label{fig6}
\end{figure}

\subsection{Numerical Examples:- L\'evy Equilibrium}

 We consider three types of bath particle velocity PDFs,
for large values of
$|v_m| \to \infty$ these PDFs exhibit 
  $f\left( \tilde{v}_m \right) \propto |\tilde{v}_m|^{-5/2}$,
which implies $\alpha = 3/ 2$.\\
(i) Case 1
\begin{equation}
f\left( \tilde{v}_m \right) = { N_1 \over \left( 1 + c \sqrt{ { m \over T} } |\tilde{v}_m|\right)^{5/2} }, 
\label{eqexs1}
\end{equation}
where the normalization constant is $ N_1 = c 3 \sqrt{ m/T} / 4$ and 
$c=3^{2/3} / 2$.  
The Fourier transform of this equation can be expressed
in terms of a Hyper-geometric function as in Eq. (\ref{EqMath1}).\\
(ii) 
 Case 2, 
\begin{equation}
f\left( \tilde{v}_m \right) = { N_2 \over \left( 1 + { m \tilde{v}_m ^2 \over 2 \tilde{T} }\right)^{5/4} }
\label{eqEN01}
\end{equation} 
for $-\infty< \tilde{v}_m < \infty$.
The normalization constant  is 
\begin{equation}
N_2= \sqrt{ { m \over 2 \tilde{T}}}{ \Gamma\left({1 \over 4}\right) \over 4 
\sqrt{\pi} \Gamma\left({3 \over 4}\right) }, 
\label{eqEN02}
\end{equation}
and
\begin{equation}
\tilde{T} = { 2 \over \pi^{2/3} } \left[ { \Gamma\left( { 3 \over 4}\right) \over \Gamma\left( {1 \over 4} \right) } \right]^{4/3} T.
\label{eqEN03}
\end{equation}
The bath particle characteristic function is
\begin{equation}
\bar{f} \left( k \right) = { 2^{1/4} N  \sqrt{\pi} \over \Gamma\left( 5/4\right) \left( { m \over 2 \tilde{T} } \right)^{7/8} } |k|^{3/4}  K_{3/4} \left( {k \sqrt{ 2 \tilde{T} } \over \sqrt{m} } \right),  
\label{eqEN04}
\end{equation}
which yields the small $k$ 
expansion
\begin{equation}
\bar{f} \left( k \right) = 1 -  2.34565  \left( { \tilde{T} \over m} \right)^{3/4}  |k|^{3/2} + 2 {\tilde{T} \over m } k^2 + \cdots ,
\label{eqEN05}
\end{equation}
or using Eq.
(\ref{eqEN03})
$\bar{f} \left( k \right) = 1 - \left(  T / m  \right)^{3/4}  |k|^{3/2}/\Gamma(5/2) + \cdots $. \\
(iii) Case 3, the bath particle velocity PDF is a 
L\'evy PDF with index $3/2$, whose characteristic function
is
\begin{equation}
\bar{f}\left( k \right) = \exp\left[ - \left( { T \over m} \right)^{3/4} {  |k|^{3/2} \over \Gamma\left( 5/2 \right) } \right].
\label{eqLLVV}
\end{equation}

 According to our theory these power law velocity PDFs,
yield a L\'evy equilibrium for the tracer particle,
Eq. (\ref{eqSca100}). In Fig. \ref{fig6} we show numerically
exact solution of the problem for cases (1-3). These
solutions show a good agreement between numerical results
and the asymptotic theory. The L\'evy equilibrium for the
tracer particle is not sensitive to precise shape of
the velocity distribution of the bath particle, and hence like the
Maxwell--Boltzmann equilibrium is stable. 

\begin{figure}[htb]
\epsfxsize=20\baselineskip
\centerline{\vbox{
        \epsfig{file=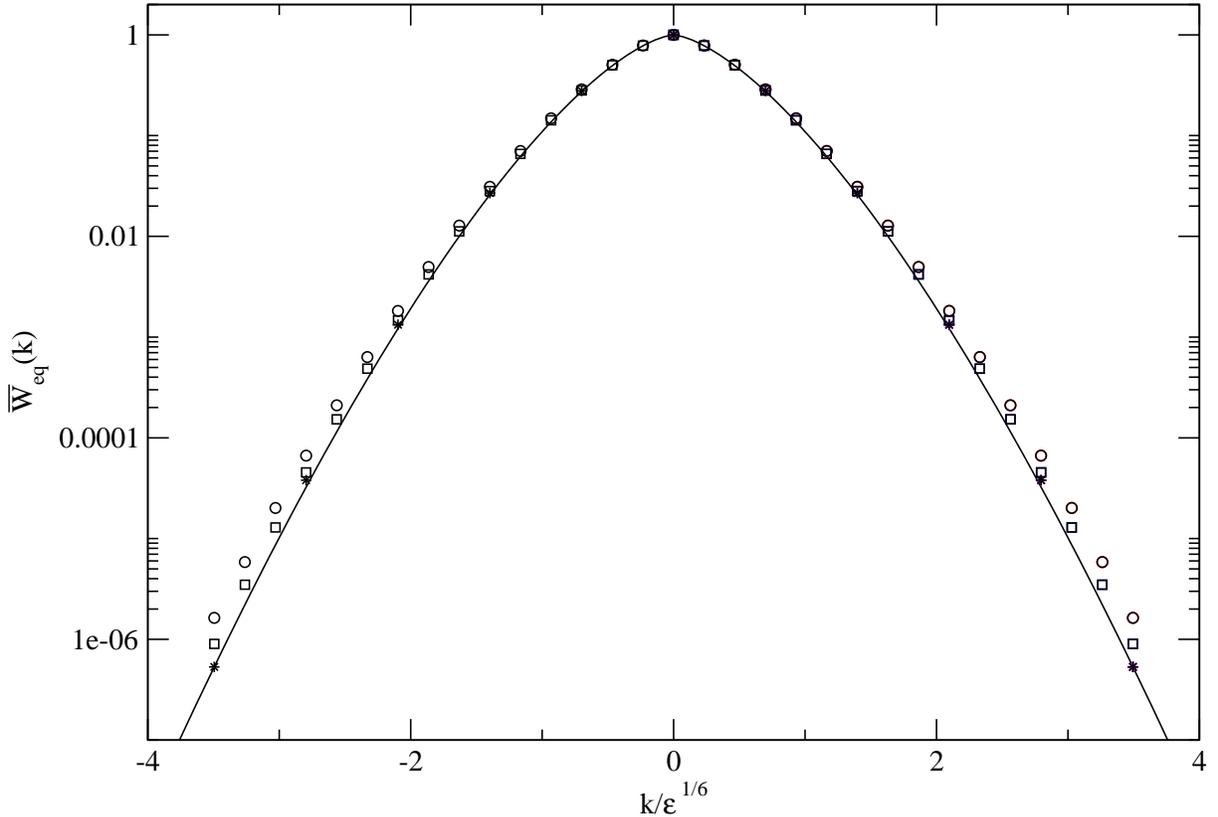, width=0.85\linewidth, angle=-90}  }}
\caption {
 The tracer particle equilibrium characteristic function
$\bar{W}_{eq}(k)$ versus $k$.
The velocity distribution of the bath particle is described
by Eq. (\protect{\ref{eqEN04}}).  
The solid curve is the theoretical prediction the L\'evy characteristic
function
$\bar{W}_{eq}(k)\sim\exp\left(-2.211|{ k \over \epsilon^{1/6}}|^{3/2} \right)$. 
We used $\epsilon=1e-4$ (circles), $\epsilon = 1e-5$ (squares), and
$\epsilon=1e-6$ (stars). As $\epsilon \to 0$ number of collisions
needed to reach an equilibrium becomes large and then
a L\'evy behavior is obtained.
}
\label{fig7}
\end{figure}

 A closer look at $\bar{W}_{eq}(k)$, for finite
value of $\epsilon$, reveals expected
deviations from L\'evy behavior when $k$ is large.
In Fig. \ref{fig7} we obtained
the numerical solution of the problem  using Eqs. (\ref{eq13}) and
(\ref{eqEN04}).
As in previous sections, we fixed $\epsilon$ and used 
values of $n$ which
were large enough to obtain a stationary solution
(e.g., when $\epsilon=1e-6$ we used $n=3e6$). 
 In Fig. \ref{fig7}
we show that as $\epsilon \to 0$ the asymptotic L\'evy solution
yields an excellent approximation, for a range of $k$ in which
the characteristic function $\bar{W}_{eq} (k )$ decays
over several order of magnitude.

\section{Back to Dynamics: Fractional Fokker--Planck Equations} 
\label{SecBack}

 Much work was devoted to the investigation of 
Langevin equation with L\'evy white noise 
\cite{West,Jep,Fog,Vlad,Garb}, 
or related 
fractional Fokker--Planck
equation
\cite{Chechkin,MBK}. For example the simplest case of generalized
Brownian motion, West and Seshadri \cite{West} consider
\begin{equation}
\dot{V_M} = - \gamma V_M + \eta(t)
\label{eqLan}
\end{equation}
  where $\eta(t)$ is a
white L\'evy noise. Not surprisingly
the equilibrium in this case is 
a L\'evy distribution. While such a stochastic approach might be used
to gain some insight also on equilibrium, the approach is clearly limited.  
If we do not have first principle equilibrium, one cannot impose on
the stochastic dynamics the desired stationary solution (i.e.,
relevant Einstein
relation is not known).
Hence in several investigations the dissipation $\gamma$ in
Eq. (\ref{eqLan}) is treated as independent of the noise
intensity $\eta(t)$. In the context of collision models,
this is unphysical, since the collisions are responsible both
for the dissipation and the fluctuations. 
These problems are treated now.
Note that fractional
Fokker--Planck equations compatible with Boltzmann equilibrium
are considered in \cite{BS,BPrl,Soko}.

 We consider the fractional Fokker--Planck equation
describing the dynamics of the collision model under investigation.
We rewrite
the kinetic equation
(\ref{eq05})
\begin{equation}
{ \partial \bar{W}(k,t) \over \partial  t} = - R \bar{W}(k,t) \left[ 1 - \bar{f}\left( k \xi_2\right) \right]  +
R \left[ \bar{W} \left( k \xi_1 , t \right) - \bar{W}\left( k , t \right) \right] \bar{f} \left(k \xi_2  \right).
\label{FFeq01}
\end{equation}
To obtain the continuum approximation
we consider the weak collision limit $\epsilon \to 0$.
In this limit $\xi_1 \to 1 - 2 \epsilon$ and hence
\begin{equation}
\bar{W}\left( \xi_1 k , t \right) - \bar{W}\left( k , t \right) = - 2 \epsilon k {\partial \over \partial k } \bar{W} \left( k , t \right).
\label{FFeq02}
\end{equation}
Using
Eq.
(\ref{eq02}) 
 $\xi_2 \sim 2 \epsilon$ 
and 
$\bar{f}(k) \sim 1-|k|^{\alpha} A_{\alpha}/ \Gamma(1 + \alpha)$ when $k \to 0$ 
we find
\begin{equation}
1 - \bar{ f} \left( k \xi_2 \right) \sim { A_{\alpha} 2^{\alpha} \epsilon^{\alpha} \over \Gamma\left( 1 + \alpha \right) } |k|^{\alpha}.
\label{FFeq03}
\end{equation}
Inserting Eqs. (\ref{FFeq03}) and (\ref{FFeq02}) in Eq. (\ref{FFeq01}),
\begin{equation}
{ \partial \bar{W}(k,t) \over \partial  t} \simeq - R A_{\alpha} { \left( 2 \epsilon \right)^{\alpha} \over \Gamma\left( 1 + \alpha\right)} |k|^{\alpha} \bar{W}(k, t )  - R 2 \epsilon k {\partial \over \partial k} \bar{W}\left( k, t \right).
\label{FFeq03ab}
\end{equation}
In the next two subsections we include in  
the limiting procedure $\epsilon \to 0$ also the dependence of
$A_{\alpha}$ on mass of gas particles, using the thermodynamic and
the scaling approaches.

\subsection{Thermodynamic Approach}

Using Eq.  
(\ref{eqL04}) 
and
Eq. 
(\ref{FFeq03ab})
 we obtain 
\begin{equation}
{ \partial \bar{W}(k,t) \over \partial  t} = - D_{th} |k|^{\alpha} \bar{W}\left( k , t \right) - \gamma k {\partial  \bar{W}(k,t) \over \partial k}.
\label{FFeq04}
\end{equation}
The first term describes fractional diffusion in velocity
space, while the second term describes dissipation. The damping
$\gamma$
is related to the parameters of the
model according to
\begin{equation}
\gamma = \lim_{\epsilon \to 0, R \to \infty} 2 \epsilon R,
\label{FFeq05}
\end{equation}
where the limit is taken in such a way that $\gamma$ is finite.
Note that the limit $R \to \infty$ is used also
in standard derivations of Brownian motion, this limit
means that collision although weak are very frequent. 
The diffusion coefficient is related to the damping term
according to a generalized Einstein relation
\begin{equation}
D_{th} = {2^{\alpha - 1}  T_\alpha  \gamma \over \Gamma\left( 1 + \alpha \right) M^{\alpha - 1} }.
\label{FFeq06}
\end{equation}
Eq. (\ref{FFeq04}) when inverted yields the fractional
Fokker--Planck equation
\begin{equation}
{ \partial W(V_M,t) \over \partial  t} =  D_{th} {\partial^{\alpha}  W\left( V_M , t \right) \over \partial |V_M|^{\alpha}}  + \gamma  {\partial  \over \partial V_M } \left[ V_M W(V_M , t )\right] .
\label{FFeq07}
\end{equation}
Where $\partial^{\alpha} / \partial |x|^{\alpha}$ is a the Riesz fractional
derivative.
The important thing to notice is that 
Eq. (\ref{FFeq07}) is valid only when the thermodynamic
approach is used, as such it is limited to bath particle velocity
distribution satisfying Eq.
(\ref{eqL04}).  

\subsection{Scaling Approach}

We now consider the energy scaling approach. We use
Eq. 
(\ref{eqSca07b}) [which yields $A_{\alpha} = ( T/m)^{\alpha/2}$] and
Eq.
(\ref{FFeq03ab}),
to obtain 
\begin{equation}
{ \partial W(V_M,t) \over \partial  t} \simeq  {D_{sc} \over \epsilon^{1 - \alpha/2}}  {\partial^{\alpha}  W\left( V_M , t \right) \over \partial |V_M|^{\alpha}}  + \gamma  {\partial  \over \partial V_M } \left[ V_M W(V_M , t )\right] .
\label{FFeq07z}
\end{equation}
Now the Einstein relation reads 
\begin{equation}
D_{sc} = {2^{\alpha - 1} \over \Gamma\left(1 + \alpha\right) } \left( { T \over M} \right)^{\alpha/2} \gamma.
\label{FFeq08}
\end{equation}
Note that for $\alpha<2$
the transport coefficient
in Eq. (\ref{FFeq07z}),
does not exist since $D_{sc}/\epsilon^{ 1 - \alpha/2} \to \infty$ as
$\epsilon \to
0$.
This means that the finite $\epsilon$ limit must be considered,
and then Eq. (\ref{FFeq07z}) is only an approximation.
This limitation does not exist for the standard case $\alpha=2$.

{\bf Remark 1}  
Eqs. (\ref{FFeq07}) and
(\ref{FFeq07z}) yield the correct equilibrium according to
Eqs. 
(\ref{eqL07},
\ref{eqSca100}) respectively.

{\bf Remark 2} In the derivation of the fractional Fokker--Planck  Eqs. 
(\ref{FFeq07}) and
(\ref{FFeq07z}), we did not include higher order terms beyond
the $|k|^{\alpha}$ term in the Fourier space expansion.
These higher order terms will yield higher order fractional
derivatives, e.g. $\partial^{2 \alpha} / \partial |V_M| ^{2 \alpha},$ 
(somewhat similar to the Kramers--Moyal expansion). It is assumed
here that in the limit $\epsilon \to 0$ these terms vanish.
From the equilibrium solution we know that at least the conditions
in Eqs. 
(\ref{eq00}, 
\ref{eq22},
\ref{eqL04},
\ref{eqL05})
 must apply for this type of truncation to be valid.

{\bf Remark 3} It is interesting to consider a phase
space approach, to the problem.
This leads to recently investigated fractional Kramers equation. 
If the tracer particle moves in a time independent
force field, $F(x)$ and us-usual we assume that collisions 
do not impact directly the position of the particle, 
we may add Newton's deterministic streaming terms to the fractional
equations of motion.
Using such an heuristic approach we obtain 
$$ { \partial W(V_M,X_M,t) \over \partial  t} + 
 V_M  {\partial W(V_M,X_M,t) \over \partial  X_M } -
 { F(X_M) \over M}  {\partial  W(V_M,X_M,t) \over \partial  V_M } 
 =  $$
\begin{equation}
 D_{th} {\partial^{\alpha}  W\left( V_M , t \right) \over \partial |V_M|^{\alpha}}  + \gamma  {\partial  \over \partial V_M } \left[ V_M W(V_M , t )\right] .
\label{FFeq07a}
\end{equation}
A similar equation was given by Lutz 
\cite{Lutz} who used the influence functional
method, assuming L\'evy stable random forces are acting on
the particle.  Note that Eq. (\ref{FFeq07a}) differs from the fractional
equation obtained by Kusnezov et al \cite{Kusn}, 
for a particle coupled to the so called chaotic bath. 
A very  different approach, based on Riemann--Liouville fractional
calculus, was used in \cite{BS}. The steady 
state solution in that case being the Boltzmann equilibrium
Steady state solution of equations like (\ref{FFeq07a}),
are not well investigated, further there is no proof that the
solutions of such equation are non-negative.
Lutz makes an interesting remark on fractional Kramers equation
of the type in Eq. (\ref{FFeq07a}): if the force is linear $F(x) = - k x$
no steady state is reached, the dissipation being too weak.

\section{Summary}

Our kinetic model illustrates the relation between 
equilibrium statistical mechanics and generalized central
limit theorems. Beyond the mathematics one has to impose
physical constraints, to obtain a clear picture of equilibrium.
 For that aim we used the scaling and thermodynamic approaches.
While these approaches represent different physical
demands, we obtained
L\'evy equilibrium for both.  
However, for the L\'evy equilibrium the two approaches lead to different
types of equilibrium Eqs.  
(\ref{eqL07},
\ref{eqSca100}), indicating that relation between power law statistical
mechanics and standard thermodynamics is not straightforward.
While for the Maxwell--Boltzmann case the two approaches
yield a unique equilibrium. As mentioned the uniqueness of
the Maxwell velocity distribution  
is related to energy and
momentum conservation in the collision events. It is doubtful
if such behaviors could be obtained by guessing some form
of generalized Maxwell--Boltzmann distribution. 
 
 The merit of the kinetic approach is that there is no
need to impose on the dynamics L\'evy or any other
type of special form of power law distributions. Instead
the kinetic approach yields classes of unique equilibrium. 
It is left for future work to see if L\'evy type of behavior
is observed in other collision models. Two obvious generalizations
are models beyond the mean field approach used here (i.e., the assumption
of uniform collision rate), and the investigation of non-linear Boltzmann
equation under the condition that the initial velocity
distribution is long tailed.   

 We have emphasized already that domain of
attraction of the L\'evy equilibrium we obtained,
is not identical to the
domain of attraction of L\'evy distributions in the 
standard problem of summation of independent identically
distributed random variable \cite{Feller}. In particular
we mentioned that not all long tailed gas particle PDFs $f(\tilde{v}_m)$
with $\alpha<2$, will yield a L\'evy equilibrium for 
the tracer particle. A case where the equilibrium is
a convolution of L\'evy and Gauss distributions was briefly
mentioned, this class of equilibrium deserves further investigation.

 Finally, we note that L\'evy
distribution of velocities of vortex elements was observed in turbulent
flows and also from numerical simulations by Min et al \cite{Min}
(see also \cite{Takayasu,Tamura,Chavanis}). 
Sota et al \cite{Jap} used a Hamiltonian ring model to describe
a self gravitating system, their numerical results show that
within a certain phase, the particles velocity distribution is
L\'evy stable.
Notice that these long range interacting systems, and non-equilibrium systems,
are usually considered beyond the domain of usual statistical mechanics.
While the above mentioned systems are not directly
related to the model under investigation, the fact that L\'evy
velocity distributions emerge under very wide conditions
indicates that their stability is not only mathematical but 
is also valid in real systems.

%

\newpage
\section{Appendix A}

 In this Appendix the solution of the equation of motion for
$\bar{W}(k,t)$ 
Eq. (\ref{eq05})
is obtained, the initial condition is
$\bar{W}(k,0)= \exp[ i k V_M(0)]$. The inverse Fourier transform of
this solution yields $W(V_M,t)$ with initial condition $W(V_M,0) =
\delta[V_M - V_M(0)]$. Such a solution is obtained in 
Eq. (\ref{eq11}),
for the special 
case when $f(\tilde{v}_m)$ is a L\'evy PDF.

 Introduce the Laplace transform 
\begin{equation}
\bar{W}\left( k , s \right) = \int_0^{\infty} \bar{W} \left( k , t \right) \exp
\left(- s t \right) {\rm d} t.
\label{eqAp01}
\end{equation}
Using Eq. (\ref{eq05}) we have
\begin{equation}
s \bar{W} \left( k , s \right) - e^{ i k V_M(0) } = - R \bar{W}\left( k , s \right) + R \bar{W} \left( k \xi_1, s\right) \bar{f} \left( k \xi_2 \right),
\label{eqAp02}
\end{equation}
this equation can be rearranged to give
\begin{equation}
\bar{W}(k,s) = { e^{ i k V_M(0)} \over R + s } + { R \over R + s} \bar{W} \left( k \xi_1, s \right) \bar{f} \left(k \xi_2  \right).
\label{eqAp03}
\end{equation}
This equation is solved using the following procedure.
Replace $k$ with $k \xi_1  $ in Eq. 
(\ref{eqAp03}) 
\begin{equation}
\bar{W} \left(k \xi_1  , s \right) = { e^{ i k \xi_1  V_M(0) } \over R + s} + 
{ R \over R + s} \bar{W} \left( k \xi_1 ^2 , s \right) \bar{f} \left(k \xi_2 \xi_1  \right). 
\label{eqAp04}
\end{equation}
Eq. (\ref{eqAp04}) may be used to eliminate $\bar{W} \left(k \xi_1  , s \right)$
from Eq. (\ref{eqAp03}), yielding
$$\bar{W} \left( k , s \right) = { e^{ i k V_M(0) } \over R + s} +$$
\begin{equation}
{ R e^{ i k \xi_1 V_M (0) }  \over \left( R + s \right)^2  } \bar{f} \left( k\xi_2  \right) + { R^2 \over \left( R + s \right)^2 } \bar{W}\left(k^2 \xi_1  , s \right) \bar{f} \left( k \xi_2 \xi_1  \right) \bar{f} \left( k \xi_2  \right).
\label{eqAp05}
\end{equation}
Replacing $k$ with $k \xi_1 ^2 $ in Eq.
(\ref{eqAp03})
\begin{equation}
\bar{W}(k \xi_1 ^2 ,s) = { e^{ i k  \xi_1 ^2   V_M(0)} \over R + s } + { R \over R + s} \bar{W} \left( k \xi_1^3 , s \right) \bar{f} \left( k \xi_2 \xi_1 ^2\right).
\label{eqAp06}
\end{equation}
Inserting Eq. (\ref{eqAp06}) in Eq. 
(\ref{eqAp05}) and rearranging
\begin{widetext}
$$ \bar{W} \left( k , s \right) = { e^{ i k V_M(0) } \over R + s } + 
{ R e^{ i k \xi_1 V_M (0) }  \over \left( R + s \right)^2 } \bar{f} \left( k \xi_2  \right) + $$ 
\begin{equation}
{ R^2 e^{ i k \xi_1 ^2 V_M(0) } \over \left( R + s \right)^3 } \bar{f} \left( k  \xi_2 \xi_1  \right) \bar{f} \left( k \xi_2  \right) + 
\left( { R \over R + s } \right)^3 \bar{W} \left( k \xi_1 ^3 , s \right) \bar{f} \left( k \xi_2 \xi_1 ^2  \right) \bar{f} \left( k \xi_2 \xi_1  \right) \bar{f} \left( k \xi_2 \right). 
\label{eqAp07}
\end{equation}
\end{widetext}
Continuing this procedure yields
\begin{equation}
 \bar{W} \left( k , s \right) =
{e^{ i k V_M (0) } \over R + s } + \sum_{n=1}^{\infty} 
{ R^n \over \left( R + s \right)^{n + 1 } } 
e^{ i k \xi_1 ^n V_M(0) } 
\Pi_{i = 1} ^n \bar{f} \left( k \xi_1 ^{n -i } \xi_2 \right).
\label{eqBp07}
\end{equation}
Inverting to the time domain, using
the inverse Laplace $s \to t$ transform yields Eq. (\ref{eq06}).
The solution Eq. (\ref{eq06}) may be verified by substitution in 
Eq. (\ref{eq05}).


\newpage

\begin{thebibliography}{99}

\bibitem{Khin} A. Y. Khintchine {\em The Mathematical Foundations
of Statistical Mechanics} Dover (New York) 1948.

\bibitem{Feller} W. Feller, {\em An Introduction to Probability Theory and its
Application} Vol. 2, Wiley (New York) 1970. 

\bibitem{Montroll} E. W. Montroll, and M. F. Shlesinger,
{\em  Maximum Entropy Formalism, Fractals, Scaling Phenomena, and $1/f$ Noise: A Tale of Tails}
 J. of Statistical Physics {\it 32} 209 (1983).

\bibitem{Grigolini}  M. Annunziato, P. Grigolini, B. J. West, 
 {\em Canonical and non-Canonical equilibrium distribution},
 Phys.
Rev. E, {\it 64} 011107 (2001).

\bibitem{Chechkin} A. Chechkin, V. Gonchar, J. Klafter, R. Metzler,
and T. Tanatarov,
{\em  Stationary States of Non Linear Oscillators Driven By
 Levy Noise}
  Chemical Physics {\it 284} 233 (2002).

\bibitem{Zannette} D. H. Zanette, and M. A. Montemurro,
{\em  Thermal measurements of stationary non-equilibrium systems:
 A test for generalized  thermostatics.}
 cond mat 0212327
(2002).

\bibitem{Abe} S. Abe, A. K. Rajagopal,
{\em  Reexamination of Gibbs 
theorem and nonuniqueness of canonical ensemble theory},
Physica A, {\it 295} 172 (2001).
%

\bibitem{Long} T. Dauxois, S. Ruffo, E. Arimondo, M. Wilkens (Eds.)
{\em Dynamics and  Thermodynamics of Systems with Long Range 
Interactions}, Springer, (2002). 

\bibitem{CohenEDG} E. G. D. Cohen, 
{\em  Statistic and Dynamics},
Physica A {\it 305}, 19 (2002).

\bibitem{Gallavotti} G. Gallavotti, 
{\em  Non-equilibrium thermodynamics?},
cond--mat 0301172 (2003).

\bibitem{Tsallis} C. Tsallis, 
 {\em Possible generalizations of Boltzmann-Gibbs statistics},
J. Stat.
Phys.  {\it 52} 479 (1988).

\bibitem{Bouch1} J.--P. Bouchaud and A. Georges,
{\em  Anomalous Diffusion in Disordered Media- Statistical Mechanics, Models,
 and Physical Applications},
 Phys. Rep.
{\it 195}, 127 (1990).

\bibitem{review} R. Metzler, and J. Klafter,
{\em  The Random Walk's Guide to Anomalous Diffusion:
a Fractional Dynamics Approach},
Phys. Rep. {\it 339} 1 (2000).

\bibitem{Zaslavsky} G. M. Zaslavsky,
{\em  Chaos fractional kinetics and anomalous transport},
 Physics Report {\it 371} 461 (2002).

\bibitem{Cohen}
F. Bardou, J. P. Bouchaud, A. aspect, and C. Cohen-Tannoudji {\em L\'evy Statistics and Laser Cooling} (Cambridge, UK 2002).

\bibitem{Barkai1} E. Barkai, R. Silbey and G. Zumofen
{\em L\'evy Distribution of Single Molecule Line
 Shape Cumulants in Glasses},
Phys. Rev. Lett. {\it 84} 5339 (2000).

\bibitem{Barkai2} E. Barkai,
{\em Fractional Fokker--Planck Equation, Solution and Application
},
Phys. Rev. E, {\it 63}, 046118 (2001).

\bibitem{Barkai3}
Y. Jung, E. Barkai, and  R. Silbey,
{\em Lineshape Theory and Photon Counting Statistics for Blinking Quantum Dots:
a L\'evy Walk Process
},
Chemical Physics, {\it 284} 181 (2002).

\bibitem{Cercignani} C. Cercignani,
{\em Solution of the Boltzmann Equation} 
in 
{\em Non Equilibrium Phenomena 1},
J. L. Lebowitz, and E. W. Montroll Eds, North Holland,
(Amsterdam) 1983. 


\bibitem{Ernst} M. H. Ernst, {\em Exact Solutions
of the Nonlinear Boltzmann Equation and Related
Kinetic Equations} in
{\em Non Equilibrium Phenomena 1},
J. L. Lebowitz, and E. W. Montroll Eds, North Holland,
(Amsterdam) 1983. 

\bibitem{CL} N. Chernov, and J. L. Lebowitz, {\em Dynamics of a Massive Piston in an Ideal Gas: Oscillatory Motion and Approach to Equilibrium}
cond-mat/0212638 (2002).

\bibitem{Lar} C. Mejia-Monasterio. H. Larralde, F. Leyvraz,
{\em Coupled normal heat and matter transport in a simple model system},
Phys. Rev. Lett. {\it 86} 5417 (2001).
 
%
\bibitem{Kampen} N. G. Van Kampen, {\em Stochastic Processes in Physics
and Chemistry} North Holland (Amsterdam) 1992.

\bibitem{Opp} J. Keilson, and J. E. Storer, 
{\em On Brownian Motion, Boltzmann's Equation and
 the Fokker--Planck Equation} in 
{\em Stochastic Processes in Chemical Physics: The Master 
Equation} I. Oppenheim, K. E. Shuler and G. H. Weiss (editors) 
The MIT Press (Cambridge) 1977.

\bibitem{BF}  E. Barkai and V. Fleurov,
{\em Brownian Type of Motion of a Randomly Kicked Particle
Far and Close to the Diffusion Limit},
Phys. Rev. E, {\it 52}, 1558, (1995).

\bibitem{SokKB} I. M. Sokolov, J. Klafter, and A. Blumen, {\em Fractional
Kinetics} Physics Today {\bf 55} 48 (2002).

\bibitem{Berez}
A. M. Berezhkovskii, D. J. Bicout, and G. H. Weiss,
{\em  Bhantnagar--Gross--Krook model impulsive collisions
Collision model for activated rate processes: turnover behavior of the
 rate constant},
J. of Chemical Physics, {\it 111} 11050 (1999).

\bibitem{BS}
E. Barkai and R. Silbey
{\em  Fractional Kramers Equation},
J. Phys. Chem. B, {\it 104} 3866 (2000).

\bibitem{Sinai} N. Chernov, J. L. Lebowitz, and Ya. Sinai,
{\em  Scaling Dynamics of a Massive Piston in a Cube
Filled With Ideal Gas: Exact Results},
cond-mat
0212637 (2002).

\bibitem{Math} {\em The Mathematica Book} S. Wolfram {\em Cambridge University Press}.

\bibitem{West} B. J. West, V. Seshadri,
{\em Linear-Systems with Levy Fluctuations},
Physica A, {\it 113} 203 (1982). 

\bibitem{Jep} S. Jeperson, R. Metzler, H. C. Fogedby 
{\em  Levy flights in external force fields:
 Langevin and fractional Fokker--Planck equations and their solutions},
Phys. Rev. E,
{\it 59} 2739 (1999). 

\bibitem{Fog} H. C. Fogedby,
{\em  Levy flights in random environments},
Phys. Rev. Lett. {\it 73} 2517 (1994).

\bibitem{Vlad} M. 0. Vlad, J. Ross, and F. W. Schneider,
{\em  Levy diffusion in a force field, Huber relaxation kinetics,
 and non equilibrium thermodynamics: H theorem for enhanced diffusion
 with Levy white noise},
Phys. Rev. E
{\it 62} 1743 (2000).

\bibitem{Garb} P. Garbaczewski, R. Olkiewicz, 
 {\em Ornstein--Uhlenbeck-Cauchy Process}, 
J. of Mathematical Physics,
{\em 41} 6843 (2000).

\bibitem{MBK} R. Metzler, E. Barkai and J. Klafter,
{\em Deriving fractional Fokker--Planck equations from a generalized
master equation},
Europhys.  Lett. {\it 46} 431 (1999).


\bibitem{Lutz} E. Lutz, 
{\em  Fractional Transport Equations for Levy Stable Processes},
Phys. Rev. Lett., 
{\it 86} 2208 (2001).

\bibitem{Kusn} D. Kusnesov, A. Bulgac, and G. D. Dang, 
 {\em Quantum Levy Processes and Fractional Kinetics},
Phys. Rev. Lett.,
{\em 82} 1136 (1999).

\bibitem{BPrl} R. Metzler, E. Barkai, and J. Klafter,
{\em  Anomalous Diffusion and Relaxation Close to Thermal
Equilibrium: A Fractional Fokker-Planck Equation Approach}
Phys. Rev. Lett. {\em 82}, 3563 (1999).  

\bibitem{Soko} I. M. Sokolov, J. Klafter, and A. Blumen, 
{\em Do Strange Kinetics Imply Unusual Thermodynamics?},
Physical Review E, {\it 64},021107 (2001). 


\bibitem{Min} I. A. Min, I. Mezik, and A. Leonard,
{\em Levy Stable Distributions for Velocity and velocity difference in systems
of vortex elements},
Physics of Fluids, {\it 8} 1169 (1996).


\bibitem{Tamura} K. Tamura, Y. Hidaka, Y. Yusuf, S. Kai,
{\em  Anomalous diffusion and Levy distribution of particle velocity
 in soft mode turbulence in electronconvention},
Physica A, {\it 306} 157 (2002). 
Note that further measurements are needed
to determine if the measured velocity distribution are L\'evy or some
other form of power law distribution.

\bibitem{Takayasu} H. Takayasu
{\em Stable Distribution and Levy Process in Fractal Turbulence},
Progress of Theoretical Physics, {\it 72} 471 (1984).

%
\bibitem{Chavanis} P. H. Chavanis, {\em Statistical Mechanics of 
Two--Dimensional Vortices and Stellar Systems} in \cite{Long}.

\bibitem{Jap} Y. Sota et al,
{\em  Origin of scaling and non-Gaussian velocity distribution in 
self gravitating ring model},
Phys. Rev. E, {\it 64} 056133 (2001).


%

\end{thebibliography}
\end{document}